\documentclass[
twocolumn
]{IEEEtran}
\usepackage{tikz}
\usepackage{tikz-3dplot} 
\usepackage{siunitx}
\usepackage{cite}
\usepackage{amsmath,amssymb,amsfonts}
\usepackage{algorithmic}
\usepackage{subfigure}
\usepackage{mathrsfs}
\usepackage{float}
\usepackage{psfrag}
\usepackage{booktabs}
\usepackage{url}
\usepackage{algorithm2e}
\usepackage{tabularx}
\usepackage{soul}
\usepackage{comment}
\usepackage{graphicx,color}
\usepackage{mathtools}
\usepackage{epstopdf}

\usepackage{epstopdf}
\usepackage[nolist]{acronym}
\usepackage[capitalise]{cleveref}
\Crefname{equation}{Eq.\!}{Eqs.\!}
\Crefname{figure}{Fig.\!}{Figs.\!}
\Crefname{tabular}{Tab.\!}{Tabs.\!}
\Crefname{section}{Section\!}{Sections.\!}
\usepackage{pgfplots}
\usepackage{pgfplotstable}
\usepackage{tikz}
\usetikzlibrary{spy}
\usepgfplotslibrary{patchplots}
\usepackage{listings}
\usepackage{array}
\usepackage{amsmath}

\DeclareMathOperator*{\argmaxB}{argmax}

 \pgfplotsset{compat=1.18} 
\begin{document}
\bstctlcite{IEEEexample:BSTcontrol}
	\pgfplotsset{every axis/.append style={
			line width=1pt,
			legend style={font=\large, at={(0.97,0.85)}}},
} %
\pgfplotsset{compat=1.16}

\def\scheme{\textit{\ac{SIDCo} }} 
\def\d{d} 
\def\k{k} 
 \newcommand{\topk}{{\text{Top}_{\k}}} 
 \newcommand{\randk}{{\text{Rand}_{\k}}} 

\def\N{N}  
\def\x{\mathbf x}
\def\xi [#1]{\x_{\{#1\}}}
\def\xin[#1]{{\x}_{\{#1\}}^{n}}
\def\gr{g} 
\def\g{\mathbf \gr}
\def\gin [#1]{{\g}_{#1}^{n}}
\def\G{G} 
\def\C{\mathbb{C}_{\k}} 
\def\Tk[#1]{{\mathbb{T}}_{\k}\left\{ #1 \right\} } 
\def\Ceta{\mathbb{C}_{\eta}} 
\def\sigmak{\sigma_{\k}}  
\def\p{p} 
\def\b{\beta} 
\def\a{\alpha} 
\def\loc{a}
\def\bhat{\hat{\beta}} 
\def\ahat{\hat{\alpha}} 
\def\lochat{\hat{a}}

\def\sign{\textrm{sign}}                                              
\def\erf{\textrm{erf}}
\def\erfc{\textrm{erfc}}

\def\Hzero{{{\mathcal{H}}_{0}}}
\def\Hone{{{\mathcal{H}}_{1}}}
\def\Hl{{{\mathcal{H}}_{l}}}

\def\Dzero{{{\mathcal{D}}_{0}}}
\def\Done{{{\mathcal{D}}_{1}}}

\def\testH1overH0{\begin{array}{l}
		{\mathcal{H}}_1 \\
		\gtrless \\
		{\mathcal{H}}_0
\end{array}}

\def\testH0overH1{\begin{array}{l}
		{\mathcal{H}}_0 \\
		\gtrless \\
		{\mathcal{H}}_1
\end{array}}

\def\testD1overD0{{ \begin{array}{l}
			{\mathcal{D}}_1 \\
			\gtrless \\
			{\mathcal{D}}_0
\end{array}}}

\begin{acronym}

\acro{5G-NR}{5G New Radio}
\acro{3GPP}{3rd Generation Partnership Project}
\acro{AC}{address coding}
\acro{ACF}{autocorrelation function}
\acro{ACR}{autocorrelation receiver}
\acro{ADC}{analog-to-digital converter}
\acrodef{aic}[AIC]{Analog-to-Information Converter}     
\acro{AIC}[AIC]{Akaike information criterion}
\acro{aric}[ARIC]{asymmetric restricted isometry constant}
\acro{arip}[ARIP]{asymmetric restricted isometry property}
\acro{AoI}{Age of Information}
\acro{AOA}[AoA]{Angle of Arrival}

\acro{ARQ}{automatic repeat request}
\acro{AUB}{asymptotic union bound}
\acrodef{awgn}[AWGN]{Additive White Gaussian Noise}     
\acro{AWGN}{additive white Gaussian noise}
\acro{ToA}{Time on Air}
\acro{CER}{Code Error Rate}
\acro{DtS}{direct-to-satellite}
\acro{PHY}{physical layer}
\acro{FH}{frequency hopping}
\acro{UNB}{ultra narrowband}
\acro{DCSS}{Differential chirp spread spectrum}
\acro{GEO}{geosynchronous equatorial orbit}
\acro{CSS}{Chirp Spread Spectrum}
\acro{RFID}{Radio Frequency Identification}
\acro{LoS}{line of sight}
\acro{CRC}{cyclic redundancy check}
\acro{FEC}{forward error correction}
\acro{LPWAN}{low power wide area network}
\acro{HiGFC-FH}{High GF Coding Frequency Hopping}
\acro{MDS}{Maximum Distance Separable}
\acro{RTT}{Round Trip Time}
\acro{OCW}{Operating Channel Width}
\acro{OBW}{Occupied Band Width}
\acro{MSB}{minimum separation band}
\acro{FEC}{forward error correction}
\acro{TFA}{Time/Frequency Aloha}
\acro{LPGAN}{low-power global area network}
\acro{BCH}{Bose–Chaudhuri–Hocquenghem codes}
\acro{PA}{Pure Aloha}
\acro{RA}{Random Access}
\acro{LR-FHSS}{Long Range - Frequency Hopping Spread Specturm}
\acro{HLC-FH}{Hop-level Coding Frequency hopping}
\acro{SIC}{the Successive Interference Cancellation}
\acro{PEC}{Packet Erasure Channel}
\acro{LDPC}{low-density parity-check}
\acro{TDOA}[TDoA]{time difference of arrival}
\acro{MA}{Microphone Arrays}
\acro{GCC-PHAT}{Generalized Cross-Correlation Phase Transform}
\acro{RMSE}{Root Mean Square Error}
\acro{APSK}[PSK]{asymmetric PSK} 
\acro{MCC}{Matthews correlation coefficient}

\acro{waric}[AWRICs]{asymmetric weak restricted isometry constants}
\acro{warip}[AWRIP]{asymmetric weak restricted isometry property}
\acro{BCH}{Bose, Chaudhuri, and Hocquenghem}        
\acro{BCHC}[BCHSC]{BCH based source coding}
\acro{BEP}{bit error probability}
\acro{BFC}{block fading channel}
\acro{BG}[BG]{Bernoulli-Gaussian}
\acro{BGG}{Bernoulli-Generalized Gaussian}
\acro{BPAM}{binary pulse amplitude modulation}
\acro{BPDN}{Basis Pursuit Denoising}
\acro{BPPM}{binary pulse position modulation}
\acro{BPSK}{binary phase shift keying}
\acro{BPZF}{bandpass zonal filter}
\acro{BSC}{binary symmetric channels}              
\acro{BU}[BU]{Bernoulli-uniform}
\acro{BER}{bit error rate}
\acro{BS}{base station}
\acro{AoP}{Age of Processing}

\acro{CP}{Cyclic Prefix}
\acrodef{cdf}[CDF]{cumulative distribution function}   
\acro{CDF}{cumulative distribution function}
\acrodef{c.d.f.}[CDF]{cumulative distribution function}
\acro{CCDF}{complementary cumulative distribution function}
\acrodef{ccdf}[CCDF]{complementary CDF}               
\acrodef{c.c.d.f.}[CCDF]{complementary cumulative distribution function}
\acro{CD}{cooperative diversity}

\acro{CDMA}{Code Division Multiple Access}
\acro{ch.f.}{characteristic function}
\acro{CIR}{channel impulse response}
\acro{cosamp}[CoSaMP]{compressive sampling matching pursuit}
\acro{CR}{cognitive radio}
\acro{cs}[CS]{compressed sensing}                   
\acrodef{cscapital}[CS]{Compressed sensing} 
\acrodef{CS}[CS]{compressed sensing}
\acro{CSI}{channel state information}
\acro{CCSDS}{consultative committee for space data systems}
\acro{CC}{convolutional coding}
\acro{Covid19}[COVID-19]{Coronavirus disease}

\acro{CNN}{Convolutional Neural Network}
\acro{DAA}{detect and avoid}
\acro{DAB}{digital audio broadcasting}
\acro{DCT}{discrete cosine transform}
\acro{dft}[DFT]{discrete Fourier transform}
\acro{DR}{distortion-rate}
\acro{DS}{direct sequence}
\acro{DS-SS}{direct-sequence spread-spectrum}
\acro{DTR}{differential transmitted-reference}
\acro{DVB-H}{digital video broadcasting\,--\,handheld}
\acro{DVB-T}{digital video broadcasting\,--\,terrestrial}
\acro{DL}{downlink}
\acro{DSSS}{Direct Sequence Spread Spectrum}
\acro{DFT-s-OFDM}{Discrete Fourier Transform-spread-Orthogonal Frequency Division Multiplexing}
\acro{DAS}{distributed antenna system}
\acro{DNA}{Deoxyribonucleic Acid}
\acro{MFCC}{Melfrequency Cepstral coefficients}
\acro{EC}{European Commission}
\acro{EED}[EED]{exact eigenvalues distribution}
\acro{EIRP}{Equivalent Isotropically Radiated Power}
\acro{ELP}{equivalent low-pass}
\acro{eMBB}{Enhanced Mobile Broadband}
\acro{EMF}{electric and magnetic fields}
\acro{EU}{European union}

\acro{FC}[FC]{fusion center}
\acro{FCC}{Federal Communications Commission}
\acro{FEC}{forward error correction}
\acro{FFT}{fast Fourier transform}
\acro{FH}{frequency-hopping}
\acro{FH-SS}{frequency-hopping spread-spectrum}
\acrodef{FS}{Frame synchronization}
\acro{FSsmall}[FS]{frame synchronization}  
\acro{FDMA}{Frequency Division Multiple Access}

\acro{GA}{Gaussian approximation}
\acro{GF}{Galois field }
\acro{GG}{Generalized-Gaussian}
\acro{GIC}[GIC]{generalized information criterion}
\acro{GLRT}{generalized likelihood ratio test}
\acro{GPS}{Global Positioning System}
\acro{GMSK}{Gaussian minimum shift keying}
\acro{GSMA}{Global System for Mobile communications Association}

\acro{HAP}{high altitude platform}

\acro{IDR}{information distortion-rate}
\acro{IFFT}{inverse fast Fourier transform}
\acro{iht}[IHT]{iterative hard thresholding}
\acro{i.i.d.}{independent, identically distributed}
\acro{IoT}{Internet of Things}                      
\acro{IR}{impulse radio}
\acro{lric}[LRIC]{lower restricted isometry constant}
\acro{lrict}[LRICt]{lower restricted isometry constant threshold}
\acro{ISI}{intersymbol interference}
\acro{ITU}{International Telecommunication Union}
\acro{ICNIRP}{International Commission on Non-Ionizing Radiation Protection}
\acro{IEEE}{Institute of Electrical and Electronics Engineers}
\acro{ICES}{IEEE international committee on electromagnetic safety}
\acro{IEC}{International Electrotechnical Commission}
\acro{IARC}{International Agency on Research on Cancer}
\acro{IS-95}{Interim Standard 95}
\acro{IRSA}{irregular repetition slotted ALOHA}

\acro{LEO}{low earth orbit}
\acro{LF}{likelihood function}
\acro{LLF}{log-likelihood function}
\acro{LLR}{log-likelihood ratio}
\acro{LLRT}{log-likelihood ratio test}
\acro{LOS}{Line-of-Sight}
\acro{LRT}{likelihood ratio test}
\acro{wlric}[LWRIC]{lower weak restricted isometry constant}
\acro{wlrict}[LWRICt]{LWRIC threshold}
\acro{LoRaWAN}{Low power long Range Wide Area Network}
\acro{NLOS}{non-line-of-sight}

\acro{MB}{multiband}
\acro{MC}{multicarrier}
\acro{MDS}{mixed distributed source}
\acro{MF}{matched filter}
\acro{m.g.f.}{moment generating function}
\acro{MI}{mutual information}
\acro{MIMO}{multiple-input multiple-output}
\acro{MISO}{multiple-input single-output}
\acrodef{maxs}[MJSO]{maximum joint support cardinality}                       
\acro{ML}[ML]{maximum likelihood}
\acro{MMSE}{minimum mean-square error}
\acro{MMV}{multiple measurement vectors}
\acrodef{MOS}{model order selection}
\acro{M-PSK}[${M}$-PSK]{$M$-ary phase shift keying}                       
\acro{M-APSK}[${M}$-PSK]{$M$-ary asymmetric PSK} 

\acro{M-QAM}[$M$-QAM]{$M$-ary quadrature amplitude modulation}
\acro{MRC}{maximal ratio combiner}                  
\acro{maxs}[MSO]{maximum sparsity order}                                      
\acro{M2M}{machine to machine}                                                
\acro{MUI}{multi-user interference}
\acro{mMTC}{massive Machine Type Communications}      
\acro{mm-Wave}{millimeter-wave}
\acro{MP}{mobile phone}
\acro{MPE}{maximum permissible exposure}
\acro{MAC}{media access control}
\acro{NB}{narrowband}
\acro{NBI}{narrowband interference}
\acro{NLA}{nonlinear sparse approximation}
\acro{NLOS}{Non-Line of Sight}
\acro{NTIA}{National Telecommunications and Information Administration}
\acro{NTP}{National Toxicology Program}
\acro{NHS}{National Health Service}

\acro{OC}{optimum combining}                             
\acro{OC}{optimum combining}
\acro{ODE}{operational distortion-energy}
\acro{ODR}{operational distortion-rate}
\acro{OFDM}{orthogonal frequency-division multiplexing}
\acro{omp}[OMP]{orthogonal matching pursuit}
\acro{OSMP}[OSMP]{orthogonal subspace matching pursuit}
\acro{OQAM}{offset quadrature amplitude modulation}
\acro{OQPSK}{offset QPSK}
\acro{OFDMA}{Orthogonal Frequency-division Multiple Access}
\acro{OPEX}{Operating Expenditures}
\acro{OQPSK/PM}{OQPSK with phase modulation}

\acro{PAM}{pulse amplitude modulation}
\acro{PAR}{peak-to-average ratio}
\acrodef{pdf}[PDF]{probability density function}                      
\acro{PDF}{probability density function}
\acrodef{p.d.f.}[PDF]{probability distribution function}
\acro{PDP}{power dispersion profile}
\acro{PMF}{probability mass function}                             
\acrodef{p.m.f.}[PMF]{probability mass function}
\acro{PN}{pseudo-noise}
\acro{PPM}{pulse position modulation}
\acro{PRake}{Partial Rake}
\acro{PSD}{power spectral density}
\acro{PSEP}{pairwise synchronization error probability}
\acro{PSK}{phase shift keying}
\acro{PD}{power density}
\acro{8-PSK}[$8$-PSK]{$8$-phase shift keying}
\acro{PMF}{probability mass function}
\acro{PLR}{packet loss rate}
 
\acro{FSK}{frequency shift keying}

\acro{i.i.d}{independent and identically distributed}
\acro{QAM}{Quadrature Amplitude Modulation}
\acro{QPSK}{quadrature phase shift keying}
\acro{OQPSK/PM}{OQPSK with phase modulator }

\acro{RD}[RD]{raw data}
\acro{RDL}{"random data limit"}
\acro{ric}[RIC]{restricted isometry constant}
\acro{rict}[RICt]{restricted isometry constant threshold}
\acro{rip}[RIP]{restricted isometry property}
\acro{ROC}{receiver operating characteristic}
\acro{rq}[RQ]{Raleigh quotient}
\acro{RS}[RS]{Reed-Solomon}
\acro{RSC}[RSSC]{RS based source coding}
\acro{r.v.}{random variable}                               
\acro{R.V.}{random vector}
\acro{RMS}{root mean square}
\acro{RFR}{radiofrequency radiation}
\acro{RIS}{Reconfigurable Intelligent Surface}
\acro{RNA}{RiboNucleic Acid}
\acro{SA}{the Slotted Aloha}
\acro{SCBSES}[SCBSES]{Source Compression Based Syndrome Encoding Scheme}
\acro{SCM}{sample covariance matrix}
\acro{SEP}{symbol error probability}
\acro{SG}[SG]{sparse-land Gaussian model}
\acro{SIMO}{single-input multiple-output}
\acro{SINR}{signal-to-interference plus noise ratio}
\acro{SIR}{signal-to-interference ratio}
\acro{SISO}{single-input single-output}
\acro{SMV}{single measurement vector}
\acro{SNR}[\textrm{SNR}]{signal-to-noise ratio} 
\acro{sp}[SP]{subspace pursuit}
\acro{SS}{spread spectrum}
\acro{SW}{sync word}
\acro{SAR}{specific absorption rate}
\acro{SSB}{synchronization signal block}

\acro{TH}{time-hopping}
\acro{ToA}{Time-of-Arrival}
\acro{TR}{transmitted-reference}
\acro{TW}{Tracy-Widom}
\acro{TWDT}{TW Distribution Tail}
\acro{TCM}{trellis coded modulation}
\acro{TDD}{time-division duplexing}
\acro{TDMA}{Time Division Multiple Access}

\acro{UAV}{unmanned aerial vehicle}
\acro{uric}[URIC]{upper restricted isometry constant}
\acro{urict}[URICt]{upper restricted isometry constant threshold}
\acro{UWB}{ultrawide band}
\acro{UWBcap}[UWB]{Ultrawide band}   
\acro{URLLC}{Ultra Reliable Low Latency Communications}
         
\acro{wuric}[UWRIC]{upper weak restricted isometry constant}
\acro{wurict}[UWRICt]{UWRIC threshold}                
\acro{UE}{user equipment}
\acro{UL}{uplink}

\acro{WiM}[WiM]{weigh-in-motion}
\acro{WLAN}{wireless local area network}
\acro{wm}[WM]{Wishart matrix}                               
\acroplural{wm}[WM]{Wishart matrices}
\acro{WMAN}{wireless metropolitan area network}
\acro{WPAN}{wireless personal area network}
\acro{wric}[WRIC]{weak restricted isometry constant}
\acro{wrict}[WRICt]{weak restricted isometry constant thresholds}
\acro{wrip}[WRIP]{weak restricted isometry property}
\acro{WSN}{wireless sensor network}                        
\acro{WSS}{wide-sense stationary}
\acro{WHO}{World Health Organization}
\acro{Wi-Fi}{wireless fidelity}

\acro{sss}[SpaSoSEnc]{sparse source syndrome encoding}

\acro{VLC}{visible light communication}
\acro{VPN}{virtual private network} 
\acro{RF}{radio frequency}
\acro{FSO}{free space optics}
\acro{IoST}{Internet of space things}

\acro{GSM}{Global System for Mobile Communications}
\acro{2G}{second-generation cellular network}
\acro{3G}{third-generation cellular network}
\acro{4G}{fourth-generation cellular network}
\acro{5G}{5th-generation cellular network}	
\acro{6G}{6th-generation cellular network}
\acro{gNB}{next generation node B base station}
\acro{NR}{New Radio}
\acro{UMTS}{Universal Mobile Telecommunications Service}
\acro{LTE}{Long Term Evolution}

\acro{QoS}{Quality of Service}
\end{acronym}

\author{Dengke Wang,~\IEEEmembership{Member,~IEEE}, Bodhibrata Mukhopadhyay,~\IEEEmembership{Member,~IEEE}, and Mohamed-Slim Alouini,~\IEEEmembership{Fellow,~IEEE}
\thanks{D. Wang is with School of Microelectronics and Communication Engineering, Chongqing University, China. B. Mukhopadhyay is with the Department of Electronics and Communication, Indian Institute of Technology Roorkee, Uttarakhand, 247667, India.  M. S. Alouini is with Computer, Electrical, and Mathematical Sciences and Engineering Division (CEMSE), King Abdullah University of Science and Technology (KAUST), Thuwal, 23955-6900, Kingdom of Saudi Arabia.  (email: dengke.wang@cqu.edu.cn, bodhibrata@ece.iitr,  slim.alouini@kaust.edu.sa).}}
\title{A Distributed Microphone Network for Real-Time Gunshot Detection and Localization in GNSS-Challenged Large Areas
}

\maketitle

\begin{abstract}

This paper presents a scalable distributed acoustic surveillance system for real-time gunshot detection and localization in large-area, GNSS-challenged environments. The system consists of multiple low-cost MEMS microphone array units (MAUs) deployed as distributed anchors. Each MAU performs edge-level sound-event detection using a lightweight Sound Detection and Echo Suppression (SDES) algorithm, followed by a TinyML-based classifier for gunshot identification. For detected gunshots, an enhanced Generalized Cross-Correlation (eGCC) algorithm estimates intra-array time differences of arrival (TDoAs) with improved robustness under low signal-to-noise ratio and reverberant conditions. Angle-of-arrival (AoA) estimates obtained from individual MAUs are fused through a modified SCO-EV localization framework to determine the gunshot position in a global coordinate system. A misplacement calibration (MC) algorithm compensates for platform orientation errors, improving angular consistency and localization accuracy in practical field conditions. Extensive experiments conducted over coverage areas ranging from 4,985~m$^2$ to 49,063~m$^2$ demonstrate reliable operation across varying spatial scales. The proposed framework achieves a Matthews Correlation Coefficient (MCC) of 0.70 for gunshot recognition and average localization errors below 6.5~m over 25,000~m$^2$, while maintaining errors under 15~m at larger scales. The results demonstrate the practical potential of the proposed system for distributed gunshot surveillance in GNSS-challenged large areas.


\end{abstract}

\acresetall 
\section{Introduction}

Wildlife poaching, driven largely by profit, continues to pose a severe threat to global biodiversity~\cite{mozer2023introduction}. 
The Wildlife Conservation Society has highlighted that the global trade in ivory results in the slaughter of up to 100,000 elephants in Africa~\cite{wcsElephants}. Given the scale of this illegal activity, innovative technologies such as audio sensors for gunshot detection are being explored to aid in real-time monitoring and enforcement, helping to mitigate wildlife poaching.

To address this growing threat, advanced sensor technologies have been increasingly adopted for monitoring and anti-poaching efforts. Each technology brings unique strengths and limitations, making them valuable for specific aspects of wildlife protection. For instance, radar sensors provide long-range capabilities and can estimate velocity and track targets without requiring line of sight. However, they are expensive and lack stealth due to their active operation, while also being sensitive to environmental factors such as precipitation and foliage~\cite{kumar2023survey}. Magnetic sensors are effective in detecting metal objects such as vehicles or weapons, but are restricted by their very short range~\cite{alimi2023situ}. Optical sensors excel at target identification over long distances but are costly and require an unobstructed line of sight~\cite{wu2024multi}. Infrared and thermal sensors work well for night detection but struggle in hot environments~\cite{zhang2014wildlife}. Radio frequency-based sensors offer high stealth due to their use of covered coaxial cables but require buried installations and have limited range~\cite{zhu2023coaxial}. Motion sensors can classify fence intrusions but are restricted to the physical structures to which they are attached~\cite{matata2022electrified}. Seismic sensors detect seismic waves generated by footsteps or vehicles, but their effectiveness depends strongly on environmental conditions~\cite{choudhary2023survey}. Chemical sensors can mark and track poached items but are currently limited in performance, cannot prevent poaching itself, and may be intrusive to animals~\cite{burgues2020environmental}. Lastly, animal sentinels theoretically offer vast coverage and high sensitivity but face challenges in deployment, such as power requirements and the logistics of collaring~\cite{de2021timely}.

However, preventing poaching remains challenging due to the vast, complex, and dynamic habitats these animals inhabit. To achieve an effective anti-poaching system with a long lifespan, it must be energy efficient, easy to deploy, robust, scalable, capable of covering large areas, and adhere to ethical and legal standards~\cite{kamminga2018poaching}. Acoustic sensors are particularly advantageous for such anti-poaching applications. Specifically, microelectromechanical systems (MEMS) microphones using pulse-density modulation (PDM) are energy-efficient and can be powered by solar panels due to their low current consumption of $\sim280\mu$A. These microphones are also IP57 rated, ensuring durability in harsh environmental conditions, including dust and water~\cite{infineon_datasheet}. Moreover, the small size of MEMS microphones, along with their cost-effectiveness and ease of deployment across diverse terrains, makes them a preferable choice for extensive wildlife conservation areas. Additionally, acoustic sensors can detect distinct sounds, such as gunshots, from a distance, enabling swift incident localization~\cite{chen2024deep}. Acoustic sensors can be easily connected to low-cost, powerful development boards such as the Sony Spresense~\cite{sony_spresense}. This compact development board, equipped with a multicore Cortex-M4F processor, offers high computational performance with low power consumption, making it ideal for real-time audio analysis, voice recognition, and multichannel audio processing in edge computing applications.
Finally, acoustic sensors operate passively without emitting any physical waves, transmitting information only in response to poaching incidents. This minimizes any potential impact on the lives of animals.

Although there are several commercial products for acoustic localization~\cite{soundthinking}, they are expensive and inefficient to deploy in large areas~\cite{svatos2023impulse}. In light of these challenges, this work presents a low-cost, energy-efficient, and scalable acoustic surveillance system for detecting and localizing gunshots in real-time, specifically tailored for deployment in large, complex environments such as wildlife reserves.

\subsection{Related Works}

We first focus on acoustic systems designed for the detection of gunshots. Distinguishing gunshots using the spatial likelihood function is reported in~\cite{makinen2010shooter}. However, this method may be less effective in the long term, as new types of gunshots could lead to false negatives, while other loud noises might trigger false alarms. Therefore, researchers are now paying attention to using learning-based tools to include more types of gunshots and reduce the false rate. The process begins by obtaining acoustic features from the time domain~\cite{hrabina2018gunshot} or the frequency domain of the acoustic signal, such as linear prediction coefficients (LPC)~\cite{ahmed2013improving}, zero-crossing rate (ZcR)~\cite{roma2018environmental}, or \ac{MFCC}~\cite{qureshi2022gunshots}. These features are then fed to a neural network for detection and classification. Additionally, a low-cost detection system on a Raspberry Pi using deep learning is proposed using convolutional neural networks (CNN)~\cite{morehead2019low}. However, traditional learning models are often hindered by high storage costs and power consumption for edge devices. Therefore, integrating Tiny Machine Learning (TinyML)~\cite{9166461}, a low-power machine learning technology that brings intelligence to edge devices, can significantly enhance energy efficiency and is expected to play a crucial role in future acoustic sensor systems.

 Sound source localization algorithms have been developed using various approaches, such as time of arrival (ToA) and time difference of arrival (TDoA) measurements~\cite{zou2020simple,sun2024projection}. ToA localization measures the absolute arrival time of an acoustic signal at multiple receivers, making it well-suited for high-precision applications. However, it requires highly accurate synchronization between the sound source and the receivers to function effectively. In contrast, TDoA methods are based on differences in signal arrival times at multiple receivers and are typically cost-effective, as they do not require a synchronization unit.
The authors in~\cite{sun2018solution} used the modified polar representation (MPR) for source location, enabling estimators to function without knowing whether the source is near or far. An energy-efficient method that considers the position error of sensors is proposed in~\cite{9612590}, while a fast adaptive estimator tailored to acoustic sensor networks is reported in~\cite{vakulya2011fast}. Additionally, various methods for combining time information have been proposed to better suit complex environments. For instance, in~\cite{ho2023integrating}, an approach that integrates \ac{AOA} and TDoA for a joint source and sensor localization system is presented. Beyond delay-based geometry, array-processing techniques estimate the source direction directly from the acoustic field, such as the small-aperture acoustic vector-sensor array in~\cite{shen2024fourth}. A feature common to all of these methods is that their accuracy ultimately rests on the quality of the underlying time-delay measurement, which degrades under low SNR and reverberation; this has in turn motivated dedicated time-delay estimators, including the EMD-based maximum-likelihood method of~\cite{bharathi2019time} and the more recent graph-network cross-correlation approach of~\cite{jeong2026micgraphnet}. Our work adopts the same measurement-centric viewpoint, but in a distributed, low-cost, edge-processing setting in which each unit estimates intra-array delays locally and transmits only compact angular information, trading the centralized high-precision hardware assumed in these works for scalability over large areas.

A wearable prototype system presented in~\cite{sallai2011weapon} utilizes a helmet-mounted microphone array to estimate the shooter’s location by detecting both shockwaves and muzzle waves generated by gunfire. An alternative method relying solely on shockwave detection is explored in~\cite{abiri2020bullet}. A more cost-effective approach combining gunshot detection, classification, and localization is proposed in~\cite{svatos2023impulse}. However, in typical poaching scenarios, shockwaves are often significantly attenuated due to the close proximity between the poacher and the target. Consequently, muzzle waves become a more reliable acoustic cue for detecting such events. A hybrid model for three-dimensional gunshot localization that relies solely on the muzzle-blast sound is reported in~\cite{zengin2024hybrid}, which is well suited to scenarios where the shockwave is weak or absent. In~\cite{manring2023low}, a low-cost system utilizing manually synchronized audio recorders is demonstrated, achieving a mean localization error of approximately 20 meters within a 20,000 square meter area. Despite these developments, many of the existing systems remain constrained by high computational complexity and deployment costs, limiting their feasibility for large-scale anti-poaching applications. Although the low-cost design proposed in~\cite{manring2023low} reduces expenses, its limited localization accuracy presents significant challenges for effective wide-area implementation in wildlife conservation contexts.

\subsection{Contribution}

To address the absence of a scalable, low-cost solution for wide-area gunshot detection and localization in anti-poaching applications, this work introduces a novel acoustic surveillance framework that integrates both hardware and algorithmic innovations. The key contributions of this study are as follows:

\begin{itemize}
    \item \textbf{Distributed low-cost acoustic sensing architecture:}
    We design and implement a scalable surveillance platform composed of multiple Microphone Array Units (MAUs), where each MAU integrates six PDM MEMS microphones, a Sony Spresense edge processor, and LoRa-based communication. Unlike centralized high-cost acoustic platforms, the proposed architecture performs local event processing at each MAU and transmits only compact localization-related information to a gateway, thereby reducing hardware complexity, communication burden, and deployment cost.

    \item \textbf{Two-stage edge gunshot detection pipeline:} 
    We develop an energy-aware detection pipeline that combines a lightweight Sound Detection and Echo Suppression (SDES) front-end with a TinyML-based gunshot classifier (TinyGunDetect). The SDES stage filters out most irrelevant background activity before neural inference is triggered, which reduces unnecessary computation and makes continuous on-device monitoring feasible for resource-constrained platforms.

    \item \textbf{Enhanced local TDoA estimation for acoustic events:} 
    We propose an enhanced GCC-based delay estimation method (eGCC) tailored for low-SNR and reverberant gunshot recordings. The method incorporates SNR-dependent spectral weighting, coherence-aware normalization, and power-weighted fusion across event-centered frames, improving the robustness of intra-array TDoA estimation while keeping the computation lightweight enough for edge execution.

    \item \textbf{Hybrid localization with deployment-aware calibration:} 
    We introduce a Misplacement Calibration (MC) algorithm to compensate for physical misalignment in MAU deployment caused by uneven terrain and compass inaccuracies, improving real-world accuracy and ease of setup.

    \item \textbf{Field Evaluation:} 
     We build a complete working prototype and validate it through simulations and field experiments over coverage areas ranging from approximately $5{,}000~\text{m}^2$ to $49{,}000~\text{m}^2$. The results show an MCC of 0.70 for gunshot classification, localization errors below about 6.5~m over medium-scale deployments, and errors below about 15~m over the largest tested area, demonstrating the feasibility of the proposed framework for large-area acoustic monitoring.
\end{itemize}

The remainder of this paper is organized as follows. Section \uppercase\expandafter{\romannumeral2} introduces the framework structure of the low-cost acoustic surveillance system. Section \uppercase\expandafter{\romannumeral3} presents the proposed detection method, and Section \uppercase\expandafter{\romannumeral4} describes the localization algorithm. Section \uppercase\expandafter{\romannumeral5} presents the simulation and experimental results, and Section \uppercase\expandafter{\romannumeral6} concludes the paper.

This paper uses bold lower and upper case letters/symbols to represent vectors and matrices. \( \mathbf{a}(i) \) denotes the \( i \)-th element of the vector \( \mathbf{a} \), and \( \| \mathbf{a} \| \) represents the Euclidean norm of \( \mathbf{a} \). The notation \( \text{diag}(\mathbf{a}) \) refers to a diagonal matrix with the elements of \( \mathbf{a} \) placed along its diagonal. \( \mathbf{I}_N \) is the identity matrix of size \( N \times N \), and \( \mathbf{1}_N \) is an \( N \)-dimensional column vector of ones. The symbol \( \otimes \) denotes the Kronecker product of two matrices.

\begin{figure}[t!]
        \includegraphics[width=0.485\textwidth]{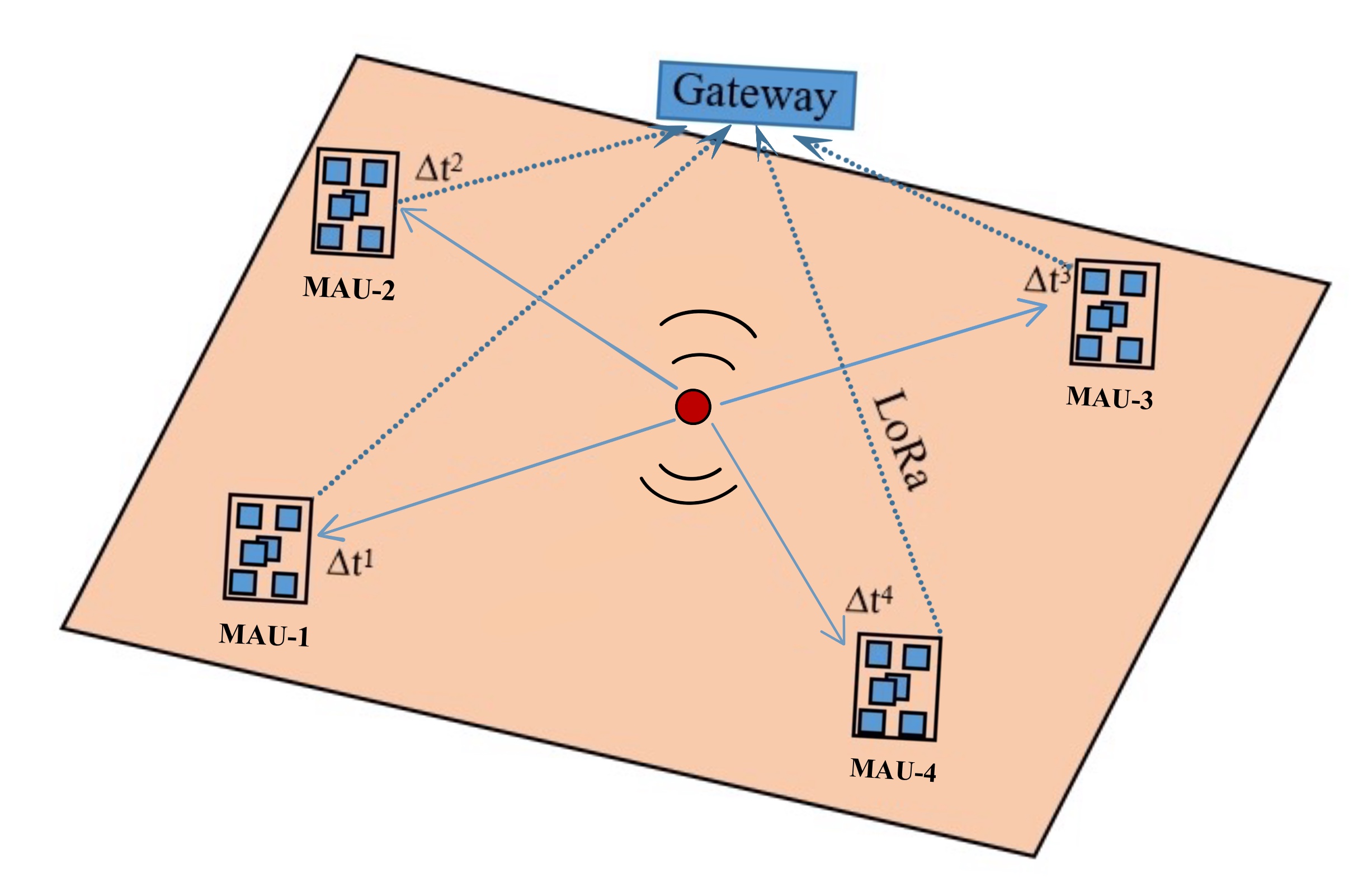}
    \caption{The proposed framework, where $\Delta t^i$ denotes the delay obtained from MAU-$i$.}
    \label{fig:framework}
\end{figure}

\section{Proposed Framework}

This paper proposes a comprehensive, real-time, and scalable acoustic surveillance system for low-cost gunshot detection and localization. As illustrated in Fig.~\ref{fig:framework}, the framework comprises at least two MAUs ($K\geq2$) and a remote server (Gateway). Each MAU comprises six microphones ($M=6$) and a microcontroller that continuously acquires surrounding audio data and performs preliminary processing. The validated multichannel signals are then processed to generate TDoA values.
The TDoA information is subsequently transmitted to the Gateway via the LoRa network. 
\begin{figure}[ht]
\centering
\subfigure[Experimental setup with MAU, where the equipped microphones are highlighted by boxes, with the red one indicating the reference microphone.]{%
    \includegraphics[width=0.8\linewidth]{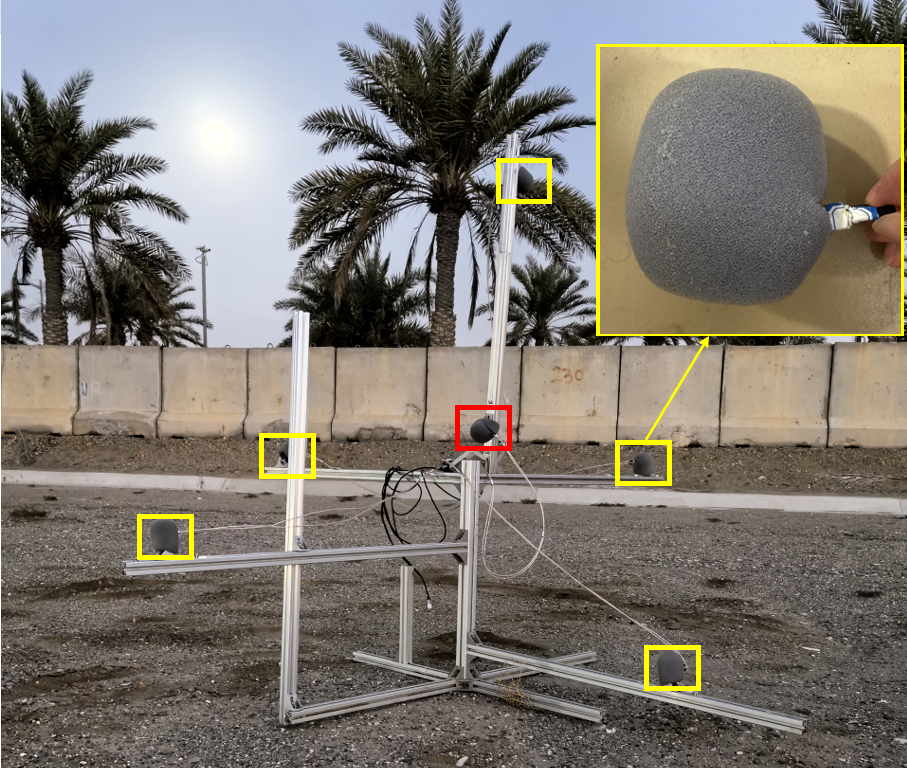}%
    \label{fig:subfig1}}

\label{fig:audio}
\vfill
\subfigure[PCB for a PDM microphone]{%
    \includegraphics[width=0.8\linewidth]{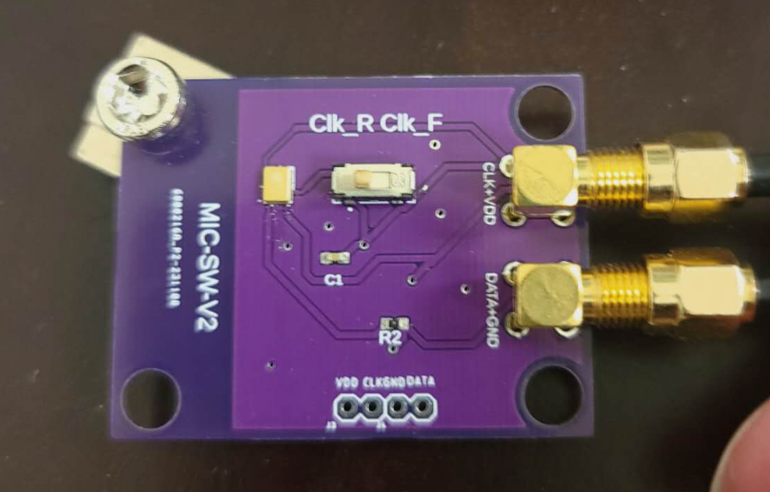}%
    \label{fig:subfig4}}
\vfill
\subfigure[Daughter board for connecting multiple microphones to the Sony Spresense board.]{%
    \includegraphics[width=0.8\linewidth]{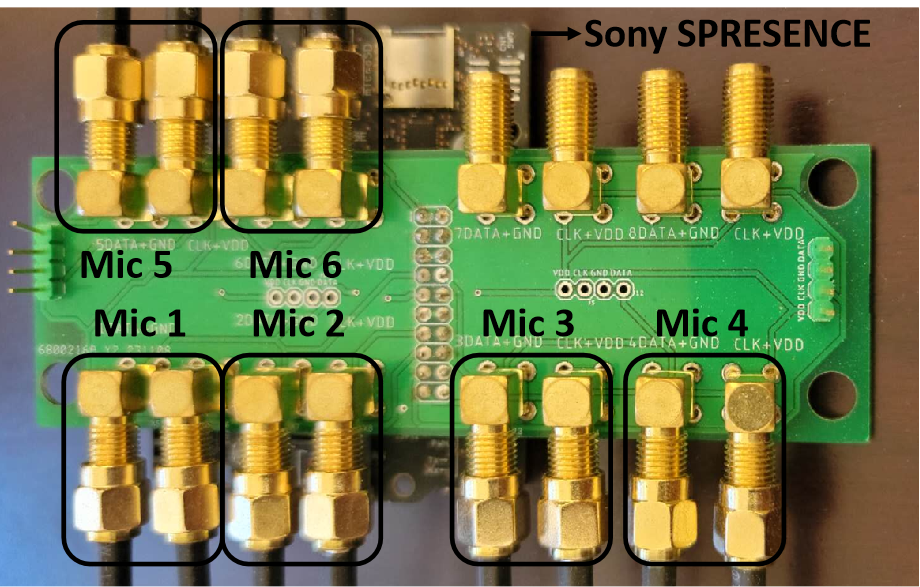}%
    \label{fig:subfig2}}
    \caption{Acoustic hardware system prototype.}

\end{figure}

\subsection{Hardware System}
To achieve the above system, we prioritize minimal hardware usage to reduce costs while ensuring high performance and reliability. The system integrates key functionalities, including position acquisition, audio recording, data processing, and transmission, through carefully selected components. 

Among these, position acquisition is essential for determining the MAU locations required by the localization algorithm and the sound-source position used as ground truth for validation. This is achieved using an RTK-GPS system powered by the multi-band u-blox ZED-F9P module~\cite{inside_gnss_2023}. Unlike ToA localization~\cite{svatos2023impulse}, which requires continuous GPS timing when the target sound is detected, our system uses GPS only during initial deployment. This setup provides the global coordinates of the speaker (acting as the shooter) and the MAU centers. A compass is then used to identify north, designated as the $Y$-axis, and to establish a local Cartesian coordinate system for each MAU.


The hardware for audio recording is shown in Fig.~2, where IM72D128V01XTMA1 microphones~\cite{infineon_datasheet} are mounted on a dedicated board for capturing audio signals. A daughterboard serves as a critical bridge, connecting the microphones to the Sony Spresense board for further processing. To ensure robust connectivity for distributed microphones placed at long distances, each microphone is linked to the daughterboard via two SMA cables. These cables are specifically designed to independently transmit time-varying data and clock signals, thereby minimizing the risk of signal degradation and crosstalk. Each microphone is covered with a sponge and positioned at a designated location on a rigid platform, as illustrated in Fig.~\ref{fig:subfig1}. This setup generates perfectly synchronized clock signals and avoids additional synchronization. 

The Sony Spresense board is used to process the audio data received from the daughterboard, leveraging its multicore architecture for real-time analysis and storage.
For metrological completeness, the relevant parameters of the sensing chain are as follows. The IM72D128 digital MEMS microphone provides a high signal-to-noise ratio, a high acoustic overload point, a flat frequency response, and an IP57 ingress rating, as specified in the datasheet~\cite{infineon_datasheet}. The audio is acquired at a sampling rate of $f_\textrm{s}=48$~kHz with a $16$-bit resolution, so that the timing quantization is $1/f_\textrm{s}=20.8\,\mu$s, corresponding to a range-difference quantization of about $7$~mm. The anchor (MAU) positions are obtained with the multi-band u-blox ZED-F9P RTK-GPS module, and the air temperature is measured to convert to the sound speed via $c\approx 331+0.606\,T_\textrm{c}$. These quantities serve as the principal metrological inputs considered in the measurement-uncertainty evaluation.
The generated TDoA values are stored on a SanDisk SD Card for local backup and simultaneously transmitted to the Gateway via the LoRa network, facilitated by the RFM95W LoRa transceiver. 

Compared with commercial acoustic localization platforms, such as acoustic camera systems or HEAD acoustics measurement arrays~\cite{GSA_47QTCA22D0036_Mod0024_060325,HEADAcoustics_HEAD_VISOR_VMA_V}, and other academic sound-localization solutions~\cite{manring2023low}, which typically integrate high-precision microphones and dedicated FPGA/DSP processors at costs ranging from several thousand to tens of thousands of US dollars, the proposed hardware architecture is more economical because it adopts low-cost digital MEMS microphones, a Spresense development board, and a lightweight LoRa module. Moreover, the modular and distributed system architecture enables straightforward network expansion, allowing additional sensing nodes to be deployed with minimal incremental cost and configuration complexity, thereby making the overall system highly scalable for large-area monitoring applications.

\subsection{Acoustic system functionality flowchart overview}

\begin{figure}[t!]
        \includegraphics[width=0.48\textwidth]{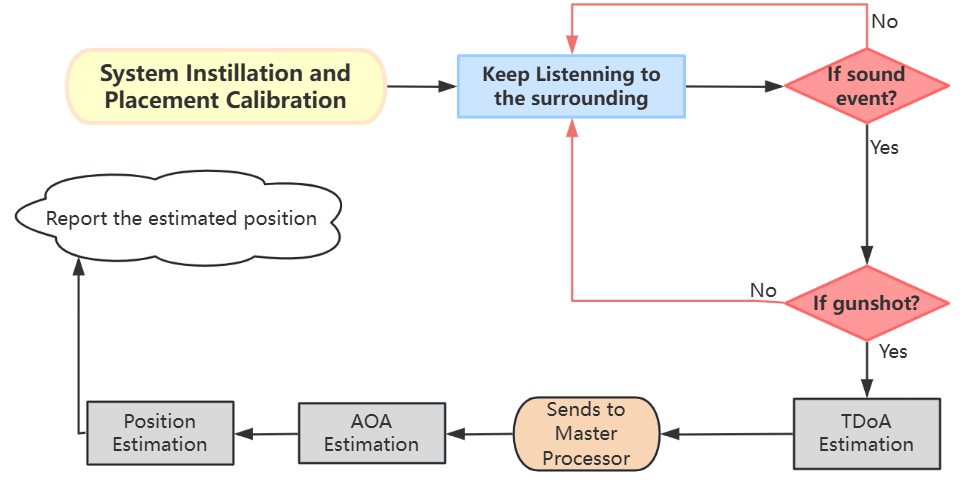}
    \caption{Acoustic system functionality flowchart.}
    \label{fig:functionality}
\end{figure}

The functionality of the acoustic system unit is shown in Fig.~\ref{fig:functionality}. The MAU locations are obtained from GPS and subsequently refined using the calibration algorithm described in Sec.~\ref{calibrationpart}. The calibrated MAU position information is then stored in the master processor for localization. Each MAU continuously monitors its surroundings and sends audio data to the sound-detection algorithm described in Sec.~\ref{presound}. Once an acoustic event is detected, the corresponding audio is passed to the gunshot classifier in Sec.~\ref{tinyml}; otherwise, the processor resumes monitoring.

If a gunshot is recognized, the local processor estimates the TDoA values using the proposed eGCC algorithm, described in Sec.~\ref{GCCa}. Upon receiving the TDoA values from the MAUs, the master processor estimates the source location using the proposed SCO-EV algorithm (see Sec.~\ref{localizationa}). Finally, the estimated position is promptly reported to local authorities.

\section{Detection algorithms}

During long-term automatic detection, acoustic events are infrequent, with most sounds being natural noise and irrelevant to the detection task. Therefore, continuously feeding all data to the gunshot detection model is energy-intensive. Moreover, false alarms from detections incur resource costs and may prevent poaching activities from being detected.

To minimize power consumption, optimize computational efficiency, and maintain high detection accuracy, we propose a two-step detection algorithm. First, the SDES algorithm detects all types of acoustic impulse events, filtering for those with distinct patterns, such as high power. Next, only the audio data that passes the initial detection stage is processed by the gunshot detection model, TinyGunDetect, trained using TinyML.

\subsection{The sound detection and echo suppression algorithm  SDES}\label{presound}

We implement a low-complexity SDES technique to filter environmental noise and suppress echoes efficiently. The detection component includes a high-pass filter and signal power criteria, while the echo suppression component focuses on regions affected by potential echoes, leveraging correlation-based methods for effective suppression. Both components are designed to minimize computational demands by operating entirely in the time domain and relying on simple operations such as squaring values, computing the median, and applying an infinite impulse response (IIR) filter.

Specifically, the original data are passed through a high-pass IIR filter with a 300-Hz cutoff to remove low-frequency noise, such as drift and hum, while preserving the higher-frequency signal components needed for analysis~\cite{lyons1997understanding}. For this continuous SDES triggering stage, we use a fourth-order Butterworth high-pass IIR filter at 300~Hz with a sampling rate of 48~kHz, implemented as a causal digital filter.\footnote{The 300~Hz cutoff was empirically selected to suppress low-frequency drift and hum while retaining the MFCC analysis range. A 300--3{,}000~Hz band-pass for impulsive-sound localization likewise uses 300~Hz as its lower bound~\cite{kotus2014detection}; nearby cutoffs and alternative filters produced no material change in our recorded data.}

\begin{table}[t!]
\centering
\caption{Pseudocode of SDES}
\label{tab:pseudocode}
\begin{tabular}{|p{0.92\columnwidth}|}
\hline
\textbf{Input:} Filtered signal, power threshold $T_\textrm{p}$, and echo threshold $T_\textrm{e}$ \\
\hline
1) Divide the signal into 50-sample segments and calculate $p_i$. \\
2) Trigger an event for a rising pair above $T_\textrm{p}$. \\
3) Re-arm the detector after a segment below $T_\textrm{p}$. \\
4) Within the next 20 segments, reject candidates correlated with the preceding rising edge above $T_\textrm{e}$. \\
5) Send accepted event windows to the gunshot classifier. \\
\hline
\textbf{Output:} Detected acoustic impulse events \\
\hline
\end{tabular}
\end{table}

After high-pass filtering, the signal is divided into 50-sample (1~ms) segments, and $p_i$ denotes the power sum of segment $i$. As summarized in Table~\ref{tab:pseudocode}, an event is triggered by a rising pair above $T_\textrm{p}=n_\textrm{p}P_\textrm{n}$, where $n_\textrm{p}$ is the empirical multiplier and $P_\textrm{n}$ is the reference noise power~\cite{svatos2019smart}. The detector is re-armed after a low-power segment; candidates within the following 20 segments are discarded when their correlation with the preceding rising edge exceeds $T_\textrm{e}$. Accepted windows are passed to the gunshot classifier.

\subsection{Gunshot Detection Algorithm}\label{tinyml}

Accurately identifying gunshots in practical scenarios is challenging due to the presence of environmental noise and reflections. Spectral analysis often captures the characteristics of the surrounding acoustic environment rather than the unique features of gunshots. Moreover, the variability in spectral characteristics caused by reflections and diffractions limits the reliability of methods like wavelet transforms for gunshot identification~\cite{svatos2023impulse}.

To address these challenges, one effective approach is the use of Cepstral coefficients, which provide a robust representation of audio features. In this work, we propose a gunshot detection model based on an artificial intelligence framework, trained using MFCC extracted from audio data.
The process for generating MFCC features follows the methodology outlined in~\cite{svatos2023impulse}, and the key parameters for extracting MFCC features are summarized in Table~\ref{tab:mfcc}. These parameters are carefully selected to optimize feature extraction for robust acoustic signal analysis. MFCC features serve as input to the classifier, where a one-dimensional convolutional neural network (CNN) is employed for accurate classification of gunshot events.\footnote{The MFCC front end separately analyzes the 300--24{,}000~Hz range through 32 filter banks. This broadband analysis range retains the gunshot-related spectral content used for classification and overlaps the frequency region retained for the subsequent eGCC-based TDoA processing; it is distinct from the low-complexity high-pass filter used only for SDES triggering.}

\begin{table}[ht]
 \caption{The parameters of MFCC}
    \centering
    \begin{tabular}{c|c}
        Parameter & Value\\
        \hline
       Length of audio data  & 1 s\\
       Length of audio frame & 0.02 s\\
       Pre-emphasis coefficient  & 0.98\\
       Frequency range & (300,\ 24000) Hz\\
       Number of filter banks  & 32\\
       Number of features &13\\
       Size of FFT &512\\
       \hline
    \end{tabular}
    \label{tab:mfcc}
\end{table}

\subsection*{1. Data collection}
To develop a robust gunshot detection system capable of accurately distinguishing gunshots from other acoustic events in real-world environments, the collected sound data are classified into two categories: gunshots and other. All sound files have a duration of 1~s and are represented at a sampling rate of $f_\textrm{s}=48{,}000$ samples/s with 16-bit resolution, i.e., $R=2$ bytes/sample. 

The other category includes a variety of sounds such as environmental noise, human activities, and non-gunshot acoustic events\footnote{The audio data for the "other" category were sourced from multiple sources, using publicly accessible databases like Freesound~\cite{freesound} and SoundBible~\cite{soundbible}.}. The gunshot class is primarily composed of AR-15 rifle recordings, including both single-shot and multi-shot cases, sourced from~\cite{fesliyan_studios}. To ensure a balanced and robust data set, the gunshot data are subjected to various data augmentation techniques~\cite{morehead2019low} to expand the dataset, reduce the risk of overfitting, and enhance the generalization capacity of the model.
Specifically, we apply the following data-augmentation techniques:
\begin{itemize}
    \item randomly shifting the audio samples within \([-0.5, 0.5]\)~s;
    \item altering the pitch by a random factor in \([0.7, 1.3]\);
    \item adjusting the speed by a factor in \([0.7, 1.3]\); and
    \item scaling the volume by a uniformly distributed random variable.
\end{itemize}

The original recordings were first assigned to the model-development partitions (76\% training and 24\% validation) before augmentation. The four augmented derivatives of each recording were generated only within the partition of their parent recording; thus, no original recording or its augmented versions appeared in both partitions. The resulting model-development dataset contained approximately 600 gunshot segments and 4,400 segments for the "other" category. After the model and post-processing configuration were fixed, final end-to-end testing was conducted on an independent set of continuous audio streams containing 138 gunshot and 140 non-gunshot events. Finally, we applied uniform weighting to all samples to further reduce the likelihood of overfitting.

\subsection*{2. Machine Learning process}

CNNs are a powerful and well-suited architecture for gunshot detection due to their ability to effectively model and extract patterns from audio features. By applying convolutional filters across small regions of the input data, CNNs can capture spatial and temporal relationships within the extracted MFCC features. These filters iteratively detect latent features across successive layers, enabling the network to distinguish subtle variations in sound properties, such as the distinct temporal patterns of gunshots. Furthermore, the hierarchical structure of CNNs ensures robust feature extraction while maintaining computational efficiency, making them ideal for real-time acoustic analysis~\cite{morehead2019low}.

\begin{figure}[!t]
\includegraphics[width=0.48\textwidth]{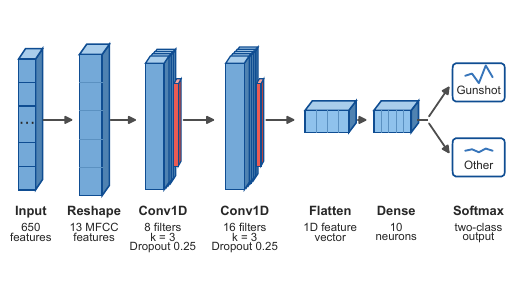}
    \caption{Architecture of the proposed CNN classifier.}
    \label{fig:CNN}
\end{figure}

Our one-dimensional CNN architecture consists of an input layer, a reshape layer, and two 1D CNN layers, each followed by a dropout layer, a flattened layer, a fully connected layer, and an output layer, as shown in Fig.~\ref{fig:CNN}. The input vector for the 1D CNN has a size of 650, corresponding to the feature amount of a 1-second audio clip. The input data is reshaped into 13 columns, each representing an MFCC feature. The first 1D CNN layer, with 8 neurons and a kernel size of 3, extracts local features from the MFCC data, capturing basic acoustic patterns like transients and frequency components common in gunshots. Dropout layers with a rate of 25\% follow each CNN layer to reduce overfitting by randomly deactivating units during training. 

The second 1D CNN layer has 16 neurons and a kernel size of 3, refining the feature extraction by capturing more complex patterns related to gunshots using a larger number of neurons. The flatten layer converts the 2D output of the CNN layers into a 1D vector, which is fed into the fully connected layer with 10 neurons. This layer combines the extracted features to make the final classification. The output layer uses softmax activation to provide a probability distribution between the "gunshot" and "other" classes.
To optimize training, we use a learning rate schedule that reduces the learning rate by a factor of 0.005 during plateaus, enabling more precise updates as the model converges.

Finally, the CNN model was trained using the Adam optimizer with a learning rate of 0.0001, a batch size of 32, and 100 epochs, employing categorical cross-entropy as the loss function. The trained model was then uploaded to the Edge Impulse platform and further optimized using quantization, pruning, weight clustering, and weight sharing to generate a low-cost TinyML model~\cite{hymel2022edge}. The optimized model was subsequently tested and validated before deployment on the target hardware system. We do not claim algorithmic novelty for the MFCC or 1D CNN components; likewise, the CNN training procedure and Edge Impulse optimization are standard deployment steps rather than methodological contributions. Instead, the classifier operates as a TinyML event gate: SDES limits inference to impulse candidates, and only recognized gunshots initiate local eGCC/TDoA processing and subsequent localization. This trigger-based integration enables continuous on-device monitoring while avoiding unnecessary processing and false localization triggers.

\section{Localization algorithms}

In this section, we introduce the proposed localization framework, which consists of three stages. First, the eGCC algorithm computes the TDoA values. Next, the grouped TDoA values from each MAU are input into the modified SCO algorithm to estimate the AoA. Finally, these estimates are combined with the corresponding MAU positions, and the EV algorithm is applied to determine the precise location of the gunshot.

\subsection{The TDoA estimation algorithm}\label{GCCa}

The proposed hardware architecture enables direct computation of time delays between microphones within each MAU, thereby reducing both synchronization errors and the energy required for data transmission. This design, however, necessitates local audio processing on the MAU itself. To ensure low-latency operation while maintaining high accuracy, we introduce the eGCC algorithm. Specifically optimized for low-resource environments, eGCC reduces the volume of data processed and minimizes computational overhead. The algorithm consists of two primary components: (1) an improved weighted GCC-PHAT technique for accurately estimating time delays between microphone pairs, and (2) a power-weighted averaging approach that consolidates TDoA values based on the signal energy of their respective audio segments, effectively reducing computational load without sacrificing estimation precision.


To address the limitations of conventional GCC-PHAT under low \ac{SNR} or heavy reverberation~\cite{kang2016pre}~\cite{wang2022optimization}, our approach combines magnitude and phase information while considering multiple parameters to balance their contributions. In contrast to the widely adopted normalization factor $\beta$, we introduce an SNR-dependent frequency weighting function $G(f)$ to emphasize contributions based on magnitude. Additionally, a locally smoothed coherence- and energy-aware reliability regularizer $R(f)$ is incorporated into the weighting function. Our proposed method reduces the worst-case SNR without errors from 4 dB to 1 dB, enabling accurate estimation in high-noise environments and over a broader range, as demonstrated in the results section.

Next, we outline the steps for generating TDoA. The audio data in the time domain are first converted to the frequency domain using an $F$-point FFT. Let \( S_0(f) \) and \( S_1(f) \) represent the Fourier transforms of two input audio signals, and let \( f = \frac{k \cdot f_s}{F} \) denote the physical frequency (in Hz) corresponding to the \( k \)-th frequency bin, where \( k \in \left\{ 1, 2, \ldots, F \right\} \). Because gunshot signals typically have high energy, we apply the Wiener gain to enhance high-SNR components. The gain is defined as follows
\begin{equation}
    G_{i}(f) = \frac{\gamma_{i}(f)}{1+\gamma_{i}(f)}, \quad i\in\{0,1\},
\end{equation}
where \( \gamma_{i}(f) \) is the signal-to-noise ratio at frequency \( f \), estimated by treating the non-event signal as noise, which can be expressed as
\begin{equation}\label{snrf}
    \gamma_{i}(f) = \frac{S_{i}^2(f)}{P_{\text{n}}(f)}, \quad i\in\{0,1\},
\end{equation}
where $P_{\text{n}}(f)$ represents the power spectral densities of the signal during the non-event period.
The weighted signal is then given by
\begin{equation}
    \hat{S}_i(f) = S_i(f) G_{i}(f)^\eta, \quad i \in\{0,1\},
\end{equation}
where \( \eta = \max(2, \Bar{\gamma}) \) is a tunable parameter that prevents over-amplification when the overall SNR is high, and $\Bar{\gamma}$ is the measured average \ac{SNR}, calculated for each frequency bin using \eqref{snrf} and then averaged geometrically across the entire frequency range.

We apply a weighting function to emphasize the phase information of the signals while reducing the influence of signal magnitudes~\cite{wang2022optimization}, which are more vulnerable to noise and reverberation. The resulting enhanced cross-power spectrum can be expressed as
\begin{equation}
     H_{01}(f) = \frac{\hat{S}_0(f)\hat{S}_1^*(f)}{W(f)},
\end{equation}
where $\hat{S}_i^*$ denotes the complex conjugate of $\hat{S}_i$, and $W(f)$ is the proposed weighting function, which is expressed as
\begin{equation}
    W(f) =  |\hat{S}_0(f)\hat{S}_1^*(f)|^\beta + R(f),
\end{equation}
where the factor $\beta$ normalizes the cross-power spectrum and avoids over-amplifying the noise spectral regions. To obtain a frequency-dependent reliability measure for a single event frame, the local cross- and auto-power spectra are estimated using a normalized nine-bin Hann kernel $h_m$ as
\begin{equation}
    \tilde{P}_{ij}(f) = \sum_{m=-4}^{4} h_m S_i(f-m)S_j^*(f-m), \quad i,j \in \{0,1\}.
\end{equation}
The corresponding magnitude-squared coherence is
\begin{equation}
    C_{01}(f) = \frac{|\tilde{P}_{01}(f)|^2}{\tilde{P}_{00}(f)\tilde{P}_{11}(f)+\epsilon},
\end{equation}
where $\epsilon$ prevents a zero denominator. The regularizer is then defined as
\begin{equation}
    R(f) = 1 + \lambda\left[1-C_{01}(f)\right]\left[1-\sqrt{G_0(f)G_1(f)}\right],
\end{equation}
where $\lambda=0.6$. Hence, $R(f)$ remains close to one for coherent or high-energy frequency bins and increases the regularization only for low-coherence, low-energy bins. Similarly, $S_i^*$ denotes the complex conjugate of $S_i$.
After applying the enhanced normalization, the cross-correlation function $r_{01}(t)$ is obtained by 
\begin{equation}
    r_{01}(t) = \mathcal{F}^{-1}\{H_{0,1}(f)\},
\end{equation}
where $\mathcal{F}^{-1}\{\}$ denotes the inverse FFT. The time delay $\tau$ is then estimated by finding the peak of $r_{01}(t)$, which is 
\begin{equation}
    \tau = \arg \max_t \left| r_{01}(t) \right|.
\end{equation}

However, the above estimation requires scanning each data frame with length $L_\textrm{s}$, which is equal to the FFT size, to ensure accuracy. For a 1-s single-channel recording, this corresponds to $f_\textrm{s}=48{,}000$ samples; with the sampling resolution $R=2$ bytes/sample, the raw record occupies $f_\textrm{s}R=96{,}000$ bytes. To reduce delay and computation load,
we note that SDES provides the power level and timestamp of the sound event, which identify the detected segment of length \( l_\textrm{s} \). Therefore, instead of computing the delay over long audio data, we focus on the sound data around the detected events. Assume the sound event is located at segment \( j \). We then focus on segments from \( j-I+1 \) to \( j+I-1 \), where \( I = L_\textrm{s}/l_\textrm{s} \).

Scanning the frames that contain the detected segment, we have $I$ different frames, each with its sound timestamp located in the \( k \)-th segment of the frame. Each pair of frames generates a TDoA value \( \tau_k \). The power-weighted average TDoA \( \hat{\tau} \) between the two microphones is expressed as
\begin{equation}
    \hat{\tau} = \frac{\sum_{k=1}^{I} \tau_k P(k)}{\sum_{k=1}^{I} P(k)},
\end{equation}
where \( P(k) = \sum_{i=0}^{I-1} p_{i+k-j} \) is the power of the \( k^{\text{th}} \) frame. This approach efficiently estimates the TDoA by focusing on relevant frames and using power-weighted averaging, improving both computational efficiency and delay estimation accuracy. Recall that each MAU is equipped with $M$ microphones, so the $i$-th MAU can generate at most $M-1$ TDoA values, denoted by $\Delta \mathbf{t}^i = [\hat{\tau}^i_1,\hat{\tau}^i_2,\dots,\hat{\tau}^i_{M-1}]$.

\subsection{Sound source position estimation}\label{localizationa}

In this part, we present the SCO-EV algorithm for estimating the sound source position using the estimated TDoA information. The method begins by estimating the AoA and an approximate distance from each MAU, incorporating calibrated placement information. Subsequently, the accurate source position is determined by combining these estimates with the precise MAU positions obtained by the RTK GPS system. This integration ensures high localization accuracy even in challenging environments.

We first formulate a measurement model using modified polar representation (MPR) to determine the coordinate position of the sound source. For example, the position of the sound source \( \mathbf{u} = [\theta, \phi, g]^T \), where \( \theta \) and \( \phi \) are the elevation and  azimuth angles, respectively, and \( g \) represents the
inverse range from the sound source to the origin, which is defined as the center of the MAU. Additionally, the positions of all \( M \) microphones are measured in their local Cartesian coordinate system, listed in Table. \ref{tab:params}. Without loss of generality, we omit the MAU index $i$ at the stage of estimating the AoA.

However, the actual positions of the microphones are subject to minor rotational deviations, represented by $(\Delta \theta,\Delta \phi)$, arising from misalignment of the local coordinate system. Taking into account the deviations, let $\mathbf{s}^o_j = [x_j,y_j,z_j]^T$, where $j=1,2,...,M$, denote the actual position of each microphone. The measured position $\mathbf{s}_j$ is expressed as
\begin{equation}\label{ss}
    \mathbf{s}_j = \mathbf{R}(\Delta \theta,\Delta \phi) \cdot \mathbf{s}^o_j,
\end{equation}
where $\mathbf{R}(\Delta \theta,\Delta \phi)$ is the combined rotation matrix, which can be expressed as
\begin{equation}
\begin{split}
    &\mathbf{R}(\Delta \theta,\Delta \phi) =\\
    &\begin{pmatrix}
\cos(\Delta \phi) \cos(\Delta \theta) & -\sin(\Delta \phi) & \cos(\Delta \phi) \sin(\Delta \theta) \\
\sin(\Delta \phi) \cos(\Delta \theta) & \cos(\Delta \phi) & \sin(\Delta \phi) \sin(\Delta \theta) \\
-\sin(\Delta \theta) & 0 & \cos(\Delta \theta)
\end{pmatrix}.
\end{split}
\end{equation}

Misplacements arise from slight measurement errors and can be reduced by calibration, which will be discussed later. We can apply the following approximations: \( \cos(\mathbf{x}) \approx 1 \), \( \sin(\mathbf{x}) \approx \mathbf{x} \), and \( \sin(\mathbf{x})\sin(\mathbf{x}) \approx 0 \) when \( x \rightarrow 0 \). Therefore, we can simplify \eqref{ss} as follows
\begin{equation}
\begin{split}
&\mathbf{s}_j = \mathbf{s}_j^o + \Delta \mathbf{s},\\
\end{split}
\end{equation}
where 
\begin{equation}\label{newe}
\Delta \mathbf{s} = \mathbf{B} \cdot \mathbf{s}_j^o,
\end{equation}
and 
\begin{equation}\label{getB}
    \mathbf{B} = [0,-\Delta \phi,\Delta \theta;\\
\Delta \phi,0,0;
-\Delta \theta,0,0].
\end{equation}

The measurement model considering random microphone-position errors has been described in~\cite{9612590}. In the present system, the correlated position error considered here is induced by the two residual rigid-platform rotations. Let $\boldsymbol{\alpha}=[\Delta\theta,\Delta\phi]^T$ and let $\mathbf{Q}_{\alpha}=\operatorname{cov}(\boldsymbol{\alpha})$ have units of rad$^2$. Linearization of \eqref{newe} gives $\Delta\mathbf{s}_j=\mathbf{J}_j\boldsymbol{\alpha}$, where
\begin{equation}\label{eq:micjac}
\mathbf{J}_j=
\begin{bmatrix}
z_j & -y_j\\
0 & x_j\\
-x_j & 0
\end{bmatrix}.
\end{equation}
The dimensionally consistent microphone-position covariance is therefore
\begin{equation}\label{scoqa}
\mathbf{J}=\begin{bmatrix}\mathbf{J}_1^T&\cdots&\mathbf{J}_M^T\end{bmatrix}^T,\qquad
\mathbf{Q}_\textrm{m}=\mathbf{J}\mathbf{Q}_{\alpha}\mathbf{J}^T,
\end{equation}
where the $(j,\ell)$ microphone block is $[\mathbf{Q}_\textrm{m}]_{j\ell}=\mathbf{J}_j\mathbf{Q}_{\alpha}\mathbf{J}_{\ell}^T$. Thus, $\mathbf{Q}_\textrm{m}$ has units of m$^2$ and retains both within-microphone and cross-microphone correlations generated by the common platform rotation. Here, $\mathbf{Q}_{\alpha}$ is obtained from the covariance of the post-calibration angular residuals. This revised problem can be solved using the SCO algorithm, as outlined in~\cite{9612590}, and can be expressed as
\begin{equation}\label{sco}
    \mathbf{u} = \text{SCO}(\mathbf{s}, \mathbf{r}, \mathbf{Q}_\textrm{r}, \mathbf{Q}_\textrm{m}),
\end{equation}
where \( \mathbf{r} = C_\textrm{s}\cdot \Delta t\) is the vector of range difference related to a reference microphone, $C_\textrm{s} \approx 331 + 0.606 T_\textrm{c} $ m/s, denotes the sound velocity, and $T_\textrm{c}$ is the temperature of the air~\cite{heydari2023scalable}. 
\( \mathbf{s}= [s_1,s_2,...,s_M] \) represent the positions of the measurement microphone. The range-difference covariance is propagated from the TDoA covariance as
\begin{equation}\label{eq:qrprop}
\mathbf{Q}_\textrm{r}=C_\textrm{s}^{2}\mathbf{Q}_{\Delta t}.
\end{equation}
For independent arrival-time errors with common standard deviation $\sigma_t$ and a shared reference microphone, $\mathbf{Q}_{\Delta t}=\sigma_t^2(\mathbf{I}_{M-1}+\mathbf{1}_{M-1}\mathbf{1}_{M-1}^T)$. Thus, $\mathbf{Q}_{\Delta t}$ has units of s$^2$ and $\mathbf{Q}_\textrm{r}$ has units of m$^2$. In the experiments, $\sigma_t$ is estimated from repeated eGCC delay residuals, while the $48$-kHz sampling quantization provides the resolution lower bound.

To address the far-field issue in source localization, we repeat the estimation process by redefining each MAU as the origin of the Cartesian coordinate system. Specifically, for each \( k \in [1, K] \), we obtain the inverse-range estimate \( \hat{g}_k=1/\hat r_k \) when the \(k\)-th MAU is regarded as the origin. Because a larger inverse range corresponds to a smaller source-to-MAU distance, the nearest-origin estimate is selected as follows:

\begin{equation}
    \eta = \argmaxB_k \hat{g}_k.
\end{equation}

The source location in the Cartesian coordinate system is then computed by:

\begin{equation}
\begin{bmatrix}
    x \\
    y \\
    z
\end{bmatrix}
=
\frac{1}{\hat{g}_\eta}
\begin{bmatrix}
    \cos(\hat{\theta}_\eta) \cdot \cos(\hat{\phi}_\eta) \\
    \cos(\hat{\theta}_\eta) \cdot \sin(\hat{\phi}_\eta) \\
    \sin(\hat{\theta}_\eta)
\end{bmatrix}
\end{equation}

This method significantly reduces errors caused by the far-field effect, where \( \hat{g} \) approaches zero and compromises estimation accuracy.

To estimate the source location accurately, the estimated AoA and distance from the \( K \geq 2 \) MAUs are combined using the AoA-EV algorithm~\cite{9040635}. The sound source location \(\mathbf{\hat{u}}=[\hat{\theta}, \hat{\phi}, \hat{g}]^T\) is obtained by:

\begin{equation}
        \mathbf{\hat{u}} = EV(\mathbf{M}, \Theta, \Phi, \mathbf{Q}_\textrm{a}, \mathbf{Q}_\textrm{s}),
\end{equation}

where $\mathbf{M}=[\mathbf{m}^1;\mathbf{m}^2;...,\mathbf{m}^{K}]$ is a vector of the 3D positions of the centers of each MAU, and $\mathbf{Q}_\textrm{s} = \sigma_\textrm{s}^2 \text{diag}(\mathbf{1}_K)\otimes\mathbf{I}_3 $ is the covariance matrix of the MAU positions, with $\sigma_\textrm{s}$ denoting the standard deviation of an MAU-coordinate error in meters. The vector $\Theta = [\theta_1,\theta_2,...,\theta_K]$ contains the estimated elevation angles, and $\Phi$ contains the azimuth angles. The angular covariance is
\begin{equation}\label{eq:qaprop}
\mathbf{Q}_\textrm{a}=\operatorname{blkdiag}(\mathbf{Q}_{\textrm{a},1},\ldots,\mathbf{Q}_{\textrm{a},K}),\qquad
\mathbf{Q}_{\textrm{a},k}=
\begin{bmatrix}
\sigma_{\theta,k}^2 & c_{\theta\phi,k}\\
c_{\theta\phi,k} & \sigma_{\phi,k}^2
\end{bmatrix},
\end{equation}
where $\sigma_{\theta,k}$ and $\sigma_{\phi,k}$ denote the standard deviations of the post-calibration elevation and azimuth residuals and $c_{\theta\phi,k}$ is their covariance, all evaluated from repeated residual observations. Hence, $\mathbf{Q}_\textrm{a}$ has units of rad$^2$; the inverse-range estimates are not used as angular variances.


\subsection{MCU platform misplacement calibration}\label{calibrationpart}

\begin{figure}[t!]
        \includegraphics[width=0.40\textwidth]{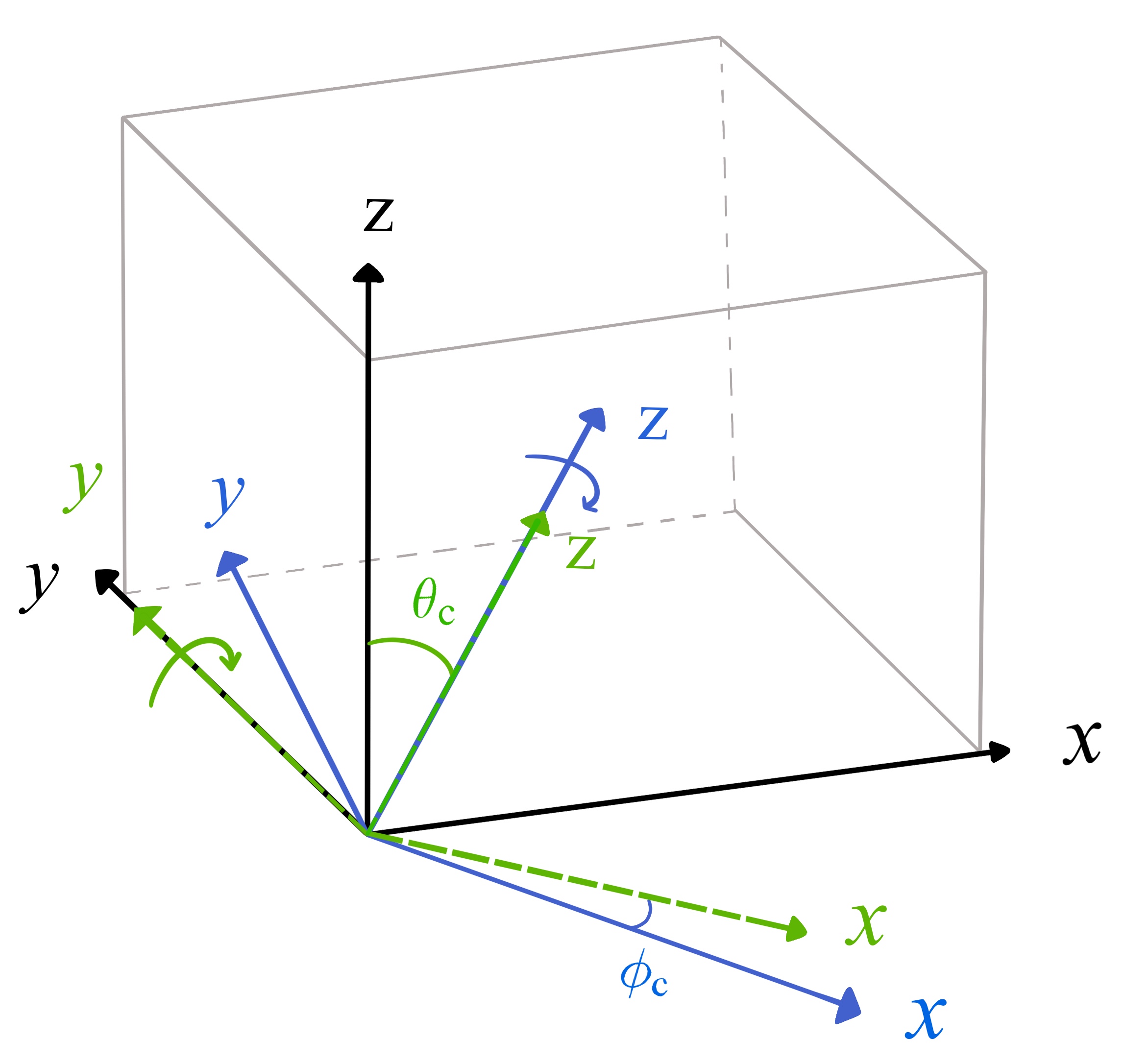}
    \caption{Illustration of the rotation of the platform. The origin state is denoted by the black one. First, the platform rotates around its y-axis by an angle $\theta_\textrm{c}$ to reach the green position, and then it rotates around its $z$-axis by an angle $\phi_\textrm{c}$ to reach the target blue position.}
    \label{fig:rotato}
\end{figure}

The MC procedure corrects only the global orientation of an MAU platform; it does not estimate independent position corrections for individual microphones. The microphone coordinates are measured once in the rigid platform coordinate system and are held fixed. After deployment, uneven ground and compass-based north alignment can introduce two small platform rotations, denoted by $(\Delta\theta,\Delta\phi)$. MC estimates the corresponding calibration angles $(\theta_\textrm{c},\phi_\textrm{c})$ from a source at a known position and applies the same rigid rotation to every measured microphone coordinate, as illustrated in Fig.~\ref{fig:rotato}.

Rotating the platform changes the positions of all microphones, so our goal is to choose two calibration angles $(\theta_\textrm{c},\phi_\textrm{c})$, which can minimize the error of AoA. In this stage,
the cost function of angle calibration is defined as
\begin{equation}
    J(\theta_\textrm{c},\phi_\textrm{c}) = |\theta_{(\theta_\textrm{c},\phi_\textrm{c})} - \theta_0 | + |\phi_{(\theta_\textrm{c},\phi_\textrm{c})} - \phi_0|,
\end{equation}
where \(\theta_0, \phi_0\) are the actual elevation and azimuth angles, which are calculated by the known calibration position of the sound source.
\(\theta_{(\theta_\textrm{c},\phi_\textrm{c})}, \phi_{(\theta_\textrm{c},\phi_\textrm{c})}\) are the estimated angles after calibration rotation. The estimated values can be obtained from~\eqref{sco} by setting the estimated noise level of the microphone position to be zero, and the rotated microphone position is
\begin{equation}
    \mathbf{s}_c = \mathbf{R}(\theta_\textrm{c},\phi_\textrm{c}) \cdot \mathbf{s},
\end{equation}
where $\mathbf{s}$ is the measured microphone position.

\begin{table}[ht]
\centering
\caption{Implementation steps of the MC algorithm}
\label{anglec}
\begin{tabular}{|l|}
\hline
\textbf{Input:} The origin point $(\theta_\textrm{c},\phi_\textrm{c}) = (0,0)$, learning rate $\alpha=0.01$,\\ the threshold $T_0= 10^{-4}$ and the maximum number of iterations\\ $N_{\textrm{max}}=100$; \\
\hline
\begin{lstlisting}[mathescape=true]
1: for each iteration $k = 1, 2, \dots, N_\text{max}$, do
2:   Calculate the current cost: $J(\theta_\textrm{c},\phi_\textrm{c})$
3:   Compute the gradient $\nabla J((\theta_\textrm{c},\phi_\textrm{c})^k)$
4:   Update: $(\theta_\textrm{c},\phi_\textrm{c})^{(k+1)} = (\theta_\textrm{c},\phi_\textrm{c})^{(k)} - \alpha \nabla J((\theta_\textrm{c},\phi_\textrm{c})^{k})$
5:   Check for convergence:
6:        if $\|(\theta_\textrm{c},\phi_\textrm{c})^{(k+1)} - (\theta_\textrm{c},\phi_\textrm{c})^{(k)}\| < T_0$
7:          Return the optimized angles $(\theta_\textrm{c},\phi_\textrm{c})^{(k+1)}$
8:        end if
9: end for
\end{lstlisting} \\
\hline
\textbf{Output:} The optimized angles $(\theta_\textrm{c},\phi_\textrm{c})^{(k+1)}$. \\
\hline
\end{tabular}
\end{table}

We use gradient descent to determine the calibration angles before localization.
The main steps of the optimization algorithm are listed in Table~\ref{anglec}. The calibrated microphone coordinates are stored in the master processor and subsequently used for localization. After applying the calibration, repeated observations provide residual pairs $\boldsymbol{\epsilon}_\ell=[\epsilon_{\theta,\ell},\epsilon_{\phi,\ell}]^T$. Their sample covariance, $\mathbf{Q}_{\alpha}=(L-1)^{-1}\sum_{\ell=1}^{L}(\boldsymbol{\epsilon}_\ell-\bar{\boldsymbol{\epsilon}})(\boldsymbol{\epsilon}_\ell-\bar{\boldsymbol{\epsilon}})^T$, is propagated to $\mathbf{Q}_\textrm{m}$ through \eqref{scoqa} to account for residual rotation-induced microphone-position errors in the SCO estimator.


\section{Results and discussion}

\begin{figure}[t!]
    \centering  
    \includegraphics[width=0.45\textwidth]{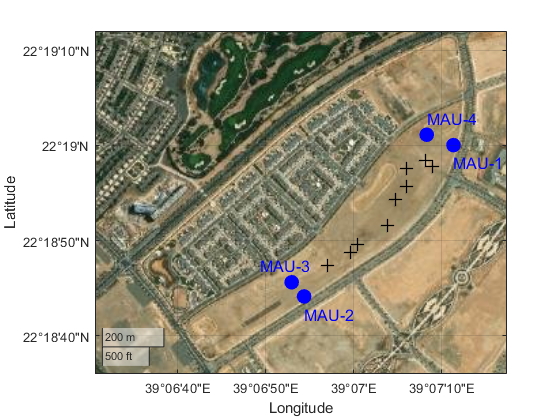}  
    \caption{Satellite map of the test site, where blue circles denote the MAU, and the black point denotes true gunshot nodes.}
    \label{fig:satellite}  
\end{figure}

\begin{figure}[t!]
	\centering
        \includegraphics[width=0.4\textwidth]{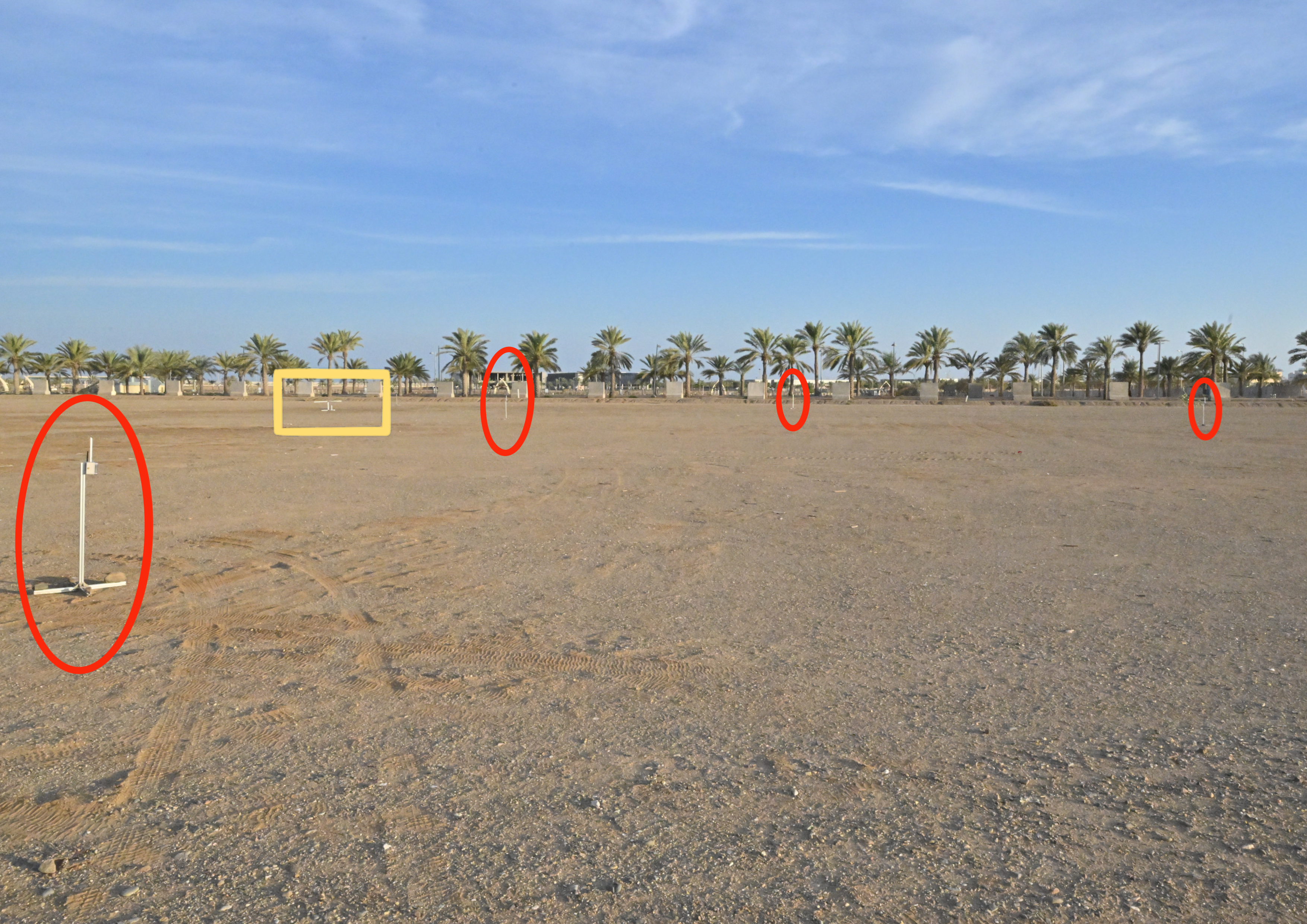}
    \caption{Experimental area, where the red circle denotes the support used to place the speaker and the yellow circle denotes the MAU.}
    \label{fig:testpoint}
\end{figure}

This section outlines the results obtained from both hardware experiments and software simulations of the proposed system. The findings are organized into four key categories: sound detection (event identification), gunshot recognition, TDoA estimation, and localization accuracy.  All experiments were conducted in the designated test area, shown in Fig. \ref{fig:satellite}, which illustrates the locations of test points and the MAU platform for one experiment. The high-fidelity speaker~\cite{Xinxiwang}, serving as the simulated gun, was placed at each test point, as depicted in Fig.~\ref{fig:testpoint}.\footnote{The playback experiments assess the multi-microphone acquisition and localization chain for muzzle-blast-dominated recordings under controlled excitation. They do not reproduce ballistic shockwaves, firearm directivity, or nonlinear source generation; live-fire validation is left for future work.} The system parameters are summarized in Table~\ref{tab:params}.

The results are organized according to the system processing order, from the acquisition of raw acoustic signals through sound detection and gunshot recognition to TDoA, angle, and position estimation.


\begin{table}[ht!]
	\renewcommand{\arraystretch}{1.5}
	\caption{System Parameters}
	\label{tab:params}
	\centering
	\begin{tabular}{cc}
		\cline{1-2}
		\multicolumn{1}{c}{Parameters}&\multicolumn{1}{c}{Values}  \\ \cline{1-2}
  Sampling Frequency ($f_\textrm{s}$) & 48000 samples/s \\
  FFT points ($F$) & 1024\\
  Temperature ($T_\textrm{c}$)& 26 $^o$C\\
  Sampling resolution ($R$)& 2 bytes\\
  Power threshold multiplier ($n_\textrm{p}$) & 3 \\
  Echo Threshold ($T_\textrm{e}$)& 0.1\\
  Segment length ($l_\textrm{s}$)&$50$\\
  Frame length ($L_\textrm{s}$)&$1024$\\
  GCC factor ($\beta$) & 0.9\\ 
  Range difference noise power ($\sigma_\textrm{r}^2$) & $10^{-1}\textrm{m}^2$\\
  MAU position noise power ($\sigma_\textrm{s}^2$) & $10^{-2}\textrm{m}^2$\\
     \cline{1-2}
      Local coordinates of Microphones\footnotemark& Meter\\
     $\mathbf{m}^1$ & (-0.06,-0.01, 0.125)\\
     $\mathbf{m}^2$ &     (0.06, 0.015, 0.8)\\
     $\mathbf{m}^3$ &      (-0.075, 0.915, 0.045)\\
     $\mathbf{m}^4$ &      (0.85, 0.015, 0.045)\\
     $\mathbf{m}^5$ &      (-0.065, -0.673, -0.44)\\
     $\mathbf{m}^6$ &      (-0.99, 0.015, 0.15)\\
     
     \cline{1-2}

		\cline{1-2}
	\end{tabular}
\end{table}
\footnotetext{The geometry balances a sufficiently large aperture for TDoA/AoA sensitivity with short PDM clock/data interconnects, which are required to limit skew, interference, and signal-integrity degradation.}

\subsection{Sound Detection and echo suppression results}
\begin{figure}[t!]
        \includegraphics[width=0.48\textwidth]{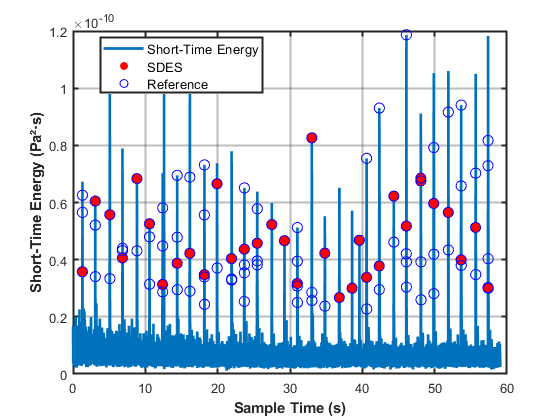}
    \caption{Results of sound detection and echo suppression are presented, where the filtered sound signal is presented by the blue curve, and the red filled circles indicate the detected sounds from the proposed SDES algorithm, while the blue circles represent those detected by the reference method.
}
    \label{fig:pred}
\end{figure}

We first present the results of the proposed SDES, benchmarked against the method described in~\cite{svatos2023impulse}. The experimental sound was recorded over a 60-second period at a distance of 248 meters. Fig.~\ref{fig:pred} illustrates the Short-Time Energy plotted against the sample time. As illustrated, both methods successfully detect all 32 sound events without omissions. However, the referenced method faces challenges in environments with heavy reverberation, leading to up to three false alarms per sound event. In contrast, the proposed SDES method achieves a substantial improvement by reducing false alarms to a single instance across all events, attaining an accuracy of approximately 96.9\%. This significant reduction in false alarms enhances detection reliability and minimizes energy consumption used in subsequent gunshot detection processes.

\subsection{Gunshot detection results}

The detected audio signals are subsequently processed by the TinyML-based gunshot detection model to classify whether the event corresponds to a gunshot. To assess the model's performance, we employ standard evaluation metrics: Accuracy (ACC), Precision (PRC), Recall (RCL), and F1-score (F1). Additionally, the Matthews Correlation Coefficient (MCC) is used to provide a comprehensive measure of the model's overall performance~\cite{manring2023low}. Let $T_{P}, T_{N}, F_{P}, F_{N}$ represent the counts of true positives, true negatives, false positives, and false negatives, respectively. The evaluation metrics are then defined as follows:
\begin{subequations}
\begin{align}
    &\textrm{ACC} = \frac{T_P+T_N}{T_P+F_P+T_N+F_N},  \\
    &\textrm{PRC} = \frac{T_P}{T_P+F_P},\\
    &\textrm{RLC} = \frac{T_P}{T_P+F_N}, \\
    &\textrm{F1} = \frac{2T_P}{2T_P+F_P+F_N}, \\
    &\textrm{MCC} = \frac{T_P T_N-F_P F_N}{\sqrt{(T_P + F_P)(T_P+F_N)(T_N+F_P)(T_N+F_N)}}.
\end{align}
\end{subequations}

After the model and post-processing configuration were fixed, final end-to-end testing was conducted on an independent set of continuous audio streams containing 138 gunshot and 140 non-gunshot events. The confusion matrix, the corresponding classification metrics, and the normalized on-device processing time are summarized in Table~\ref{gunshotdetection}. The proposed end-to-end event-recognition chain correctly identifies 113 gunshot events and correctly rejects 123 non-gunshot events. It achieves an ACC of 84.9\%, PRC of 86.9\%, RCL of 81.9\%, F1-score of 84.3\%, and MCC of 0.70. The deployed Spresense implementation reports 90~ms for MFCC/DSP processing and 45~ms for classification over a 3-s inference window; the corresponding linearly normalized values are 30~ms, 15~ms, and 45~ms in total per 1~s of audio.

\begin{table}[ht!]
\centering
\caption{Confusion matrix, classification metrics, and total normalized processing time for gunshot recognition}
\begin{tabular}{c|cc}
\textbf{Actual / Predicted} & \textbf{Gunshot} & \textbf{Non-gunshot} \\
\hline
\textbf{Gunshot} & 113 & 25 \\
\textbf{Non-gunshot} & 17 & 123 \\
\end{tabular}
\vspace{4pt}
\begin{tabular}{c|cccccc}
\textbf{Metric} & \textbf{ACC} & \textbf{PRC} & \textbf{RCL} & \textbf{F1} & \textbf{MCC} & \textbf{Proc. time} \\
\hline
\textbf{Results} & 84.9\% & 86.9\% & 81.9\% & 84.3\% & 0.70 & 45~ms \\
\hline
\end{tabular}
\label{gunshotdetection}
\end{table}

\subsection{Localization results}

Localization accuracy is evaluated across three aspects: delay, angle, and position estimation. Delay estimation is carried out through simulation, angle estimation is assessed using both simulation and experimental methods, and position estimation is evaluated based solely on experimental data.

\begin{figure}[t!]
\includegraphics[width=0.45\textwidth]{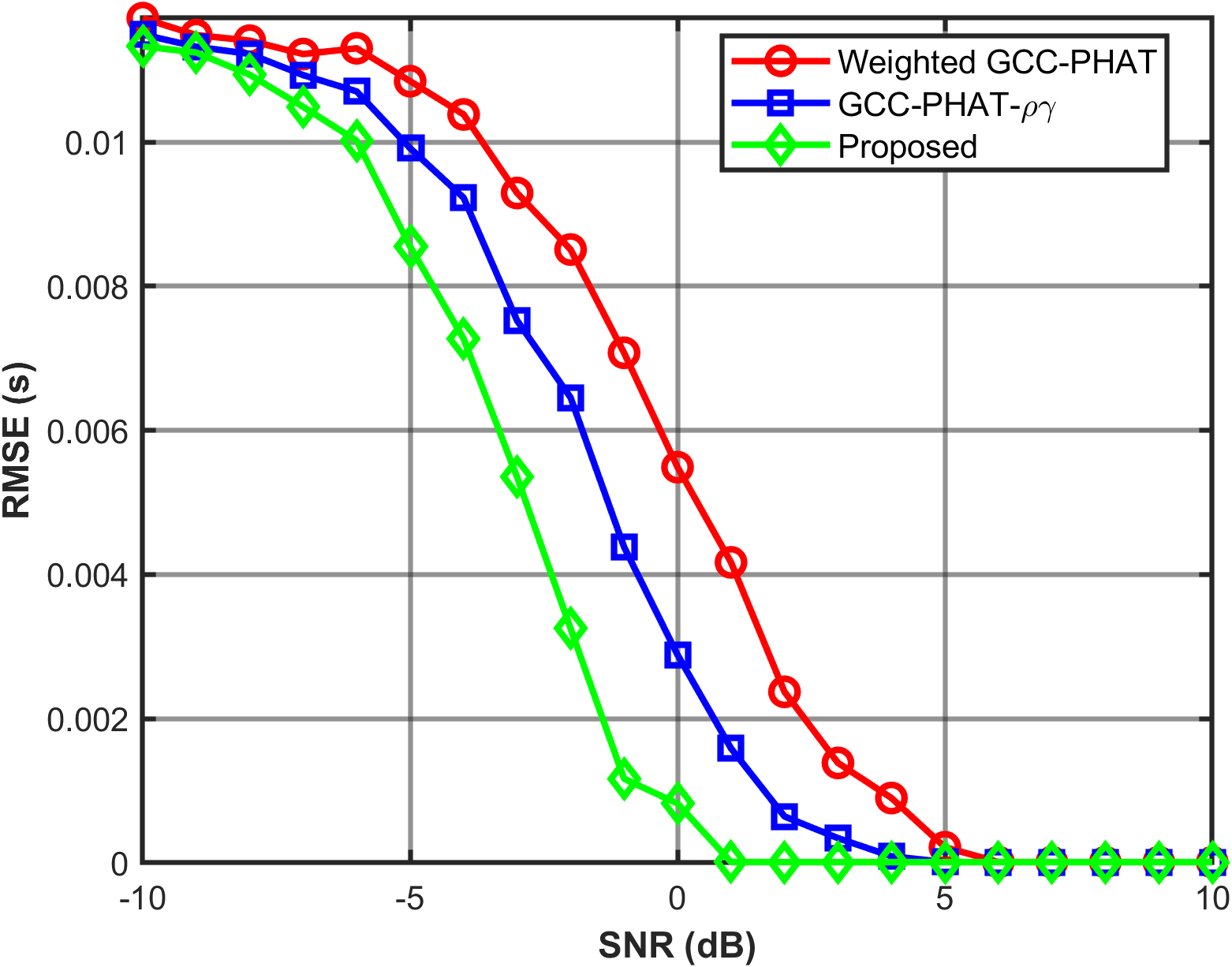}
    \caption{TDoA estimation performance as a function of SNR Levels}
    \label{fig:gcc}
\end{figure}

\begin{figure}[t]
\includegraphics[width=0.45\textwidth]{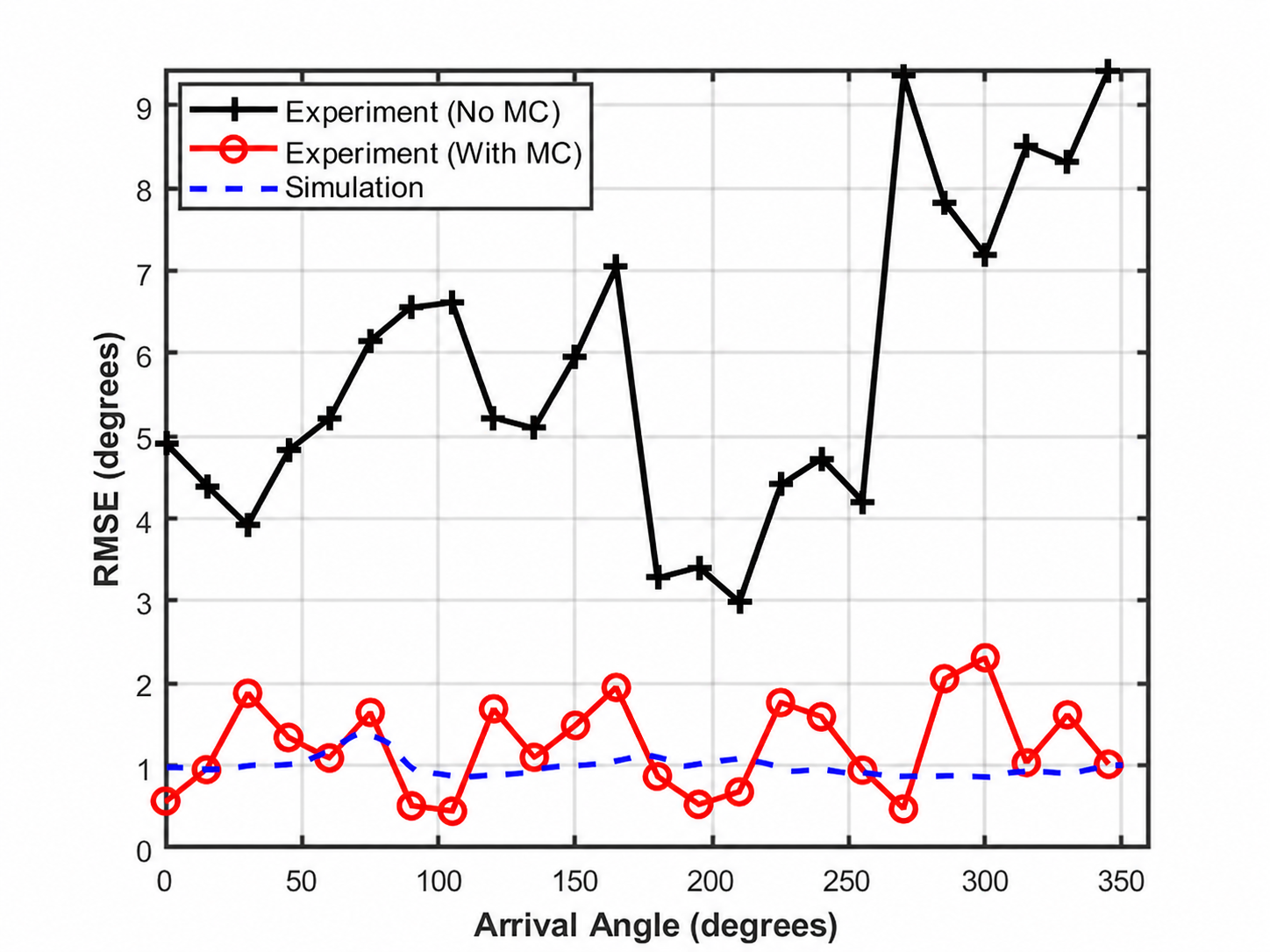}
    \caption{RMSE of angle errors when gunshots come from different directions.}
    \label{fig:angle}
\end{figure}

\begin{figure*}[ht]
\centering
\subfigure[Scenario I: Coverage area: 4985~m\textsuperscript{2}]{%
    \includegraphics[width=0.48\linewidth]{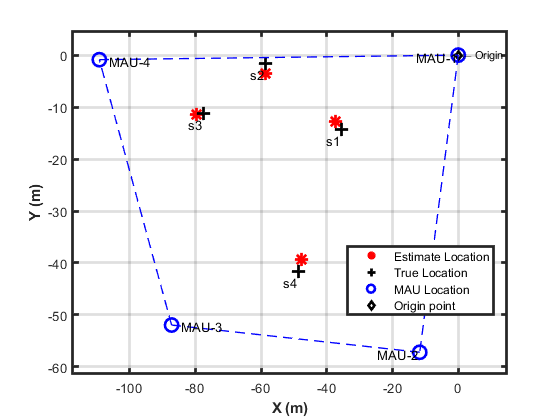}%
    \label{fig:5000}}
  \hfill
\subfigure[Scenario II: Coverage area: 16728~m\textsuperscript{2}]{%
    \includegraphics[width=0.48\linewidth]{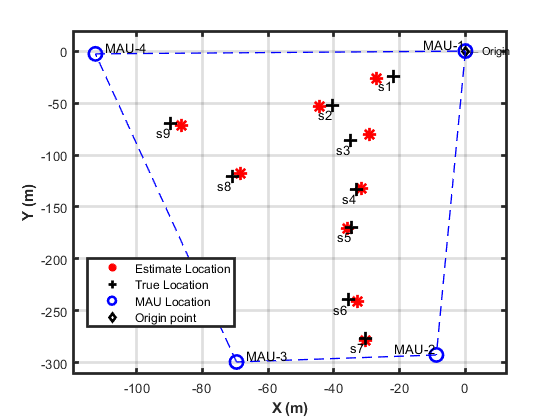}%
    \label{fig:15000}}

       \vspace{1em}
       
\subfigure[Scenario III: Coverage area: 25595~m\textsuperscript{2}]{%
    \includegraphics[width=0.48\linewidth]{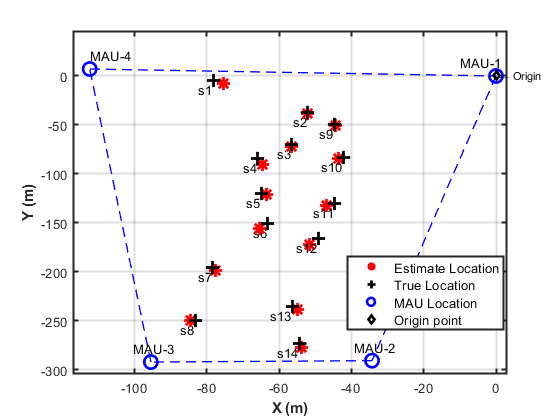}%
    \label{fig:25000}}
  \hfill
\subfigure[Scenario IV: Coverage area: 49063~m\textsuperscript{2}]{%
    \includegraphics[width=0.48\linewidth]{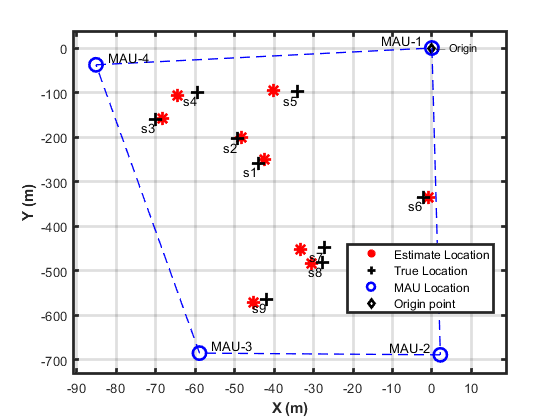}%
    \label{fig:490000}}
    \caption{Experimental sketches for different coverage areas, where the coordinate origin is defined at the MAU-1, blue circles denote the MAU, the red point denotes the estimated position, and the black point denotes true positions.}
    \label{fig:example}

\end{figure*}

\subsubsection{Delay estimation}
In this part, we present simulated TDoA-estimation results for the proposed eGCC and compare eGCC with weighted GCC-PHAT~\cite{rabinkin1996dsp} and GCC-PHAT-$\rho\gamma$~\cite{wang2022optimization}.
The simulation uses a recorded gunshot and a version delayed by 0.01~s. Both signals are corrupted with noise at varied SNR levels and processed by the compared TDoA estimators.
The estimation accuracy is quantified using the root mean square error (RMSE) of the estimated delay. As shown in Fig.~\ref{fig:gcc}, the TDoA estimation accuracy of all three methods improves as the SNR increases, while the proposed eGCC achieves the lowest RMSE across the tested SNR range. For high-accuracy estimation (RMSE \( < 2 \times 10^{-5} \) seconds), the proposed method reaches the threshold at 1 dB, whereas GCC-PHAT-\(\rho\gamma\) and Weighted GCC-PHAT reach it at 5 dB and 6 dB, respectively. For rough estimation (RMSE \( < 0.01 \) seconds), the proposed method reaches the threshold at -5 dB, matching GCC-PHAT-\(\rho\gamma\) and improving on Weighted GCC-PHAT by 2 dB. This reduced SNR threshold effectively extends the maximum detectable range of the monitoring system, since reliable delay estimation can be maintained at lower received signal power levels.

\subsubsection{Angle estimation}


To validate the microphone placement in Table~\ref{tab:params} and the gain from MC (see Sec.~\ref{calibrationpart}), we present simulation and experimental results in Fig.~\ref{fig:angle}.
We evaluated the proposed setup at different source azimuth angles. The RMSE is defined as the combined error of the estimated azimuth and elevation angles. Experimental data were collected at a distance of approximately 35~m, whereas the simulation added zero-mean Gaussian perturbations in the range-difference domain using the distance-error parameter $e_\mathrm{d}=0.01$~m. Because the implemented perturbation is $2e_\mathrm{d}n$, where $n\sim\mathcal{N}(0,1)$, its range-difference standard deviation is 0.02~m, corresponding to approximately 58~$\mu$s at a sound speed of 343~m/s; no microphone-placement perturbation was introduced.
The experimental results without the MC algorithm reveal significant angular estimation errors, ranging from 3 to 9 degrees. In contrast, the results with the MC algorithm demonstrate highly accurate and stable performance, with angular estimation errors consistently remaining below 2.5 degrees in all directions. Moreover, the experimental results with the MC algorithm exhibit average performance that closely matches the simulations. Despite potential placement variations in the physical setup, the experimental results, enhanced by the proposed algorithms, confirm a high level of accuracy in both angular and TDoA estimation.

The uneven outdoor test site caused residual platform tilt despite careful placement; mechanical aids can reduce, but not eliminate, this error. As shown in Fig.~\ref{fig:angle}, MC reduces the experimental angular RMSE from 3--9$^\circ$ to below 2.5$^\circ$, approaching the no-misalignment simulation. MC is performed once at installation, after which its calibrated coordinates are reused for localization.

To quantify the contribution of the individual components, Table~\ref{tab:mcablation} presents a sequential ablation study at a coverage area of $16{,}728~\mathrm{m}^2$. The baseline is a conventional GCC--SCO--BRPLE pipeline: conventional GCC without PHAT weighting estimates intra-array TDoAs, the original SCO estimates the AoA at each MAU, and BRPLE produces the source position. Thus, the baseline does not use EV fusion. The +eGCC and +MC configurations retain the same SCO--BRPLE localization backend, whereas the proposed configuration replaces BRPLE with EV fusion. The +eGCC configuration includes event-centered frame selection and power-weighted averaging. Weighting each delay estimate by its frame power gives greater influence to the stronger impulse frames, which typically provide a clearer direct-path correlation peak, while reducing the effect of weak frames that are more susceptible to noise and reflections. Replacing conventional GCC without PHAT weighting with eGCC reduces the localization RMSE from 15.39 to 12.04~m. Adding MC further reduces it to 6.40~m, and the complete framework with SCO-EV fusion achieves 5.83~m, corresponding to an overall reduction of 62.1\% from the baseline.

\begin{table}[t]
\centering
\caption{Component-wise ablation at a coverage area of $16{,}728~\mathrm{m}^2$}
\scriptsize
\resizebox{\columnwidth}{!}{%
\begin{tabular}{c|ccc|c}
\textbf{Configuration} & \textbf{eGCC} & \textbf{MC} & \textbf{SCO-EV} & \textbf{RMSE (m)} \\
\hline
Baseline (GCC) & $\times$ & $\times$ & $\times$ & 15.39 \\
+eGCC & $\checkmark$ & $\times$ & $\times$ & 12.04 \\
+MC & $\checkmark$ & $\checkmark$ & $\times$ & 6.40 \\
Proposed & $\checkmark$ & $\checkmark$ & $\checkmark$ & 5.83 \\
\hline
\end{tabular}
}
\label{tab:mcablation}
\end{table}

       


\subsubsection{Position estimation}

For position estimation, MAUs were positioned at the four corners, and the speaker was placed at multiple locations, as shown in Fig.~\ref{fig:satellite}. Thirty gunshot recordings were replayed at each location. The experiments were repeated with MAUs covering different areas in each trial.
Experimental sketches are shown in Fig.~\ref{fig:example}, where the blue dashed lines represent the coverage areas.\footnote{The proposed system is also capable of localizing points outside the defined coverage area.} The arena size was increased from 4,985 to 49,063~m$^2$. The estimated positions are based on the average of the computed locations.

\begin{figure}[t!]
        \includegraphics[width=0.45\textwidth]{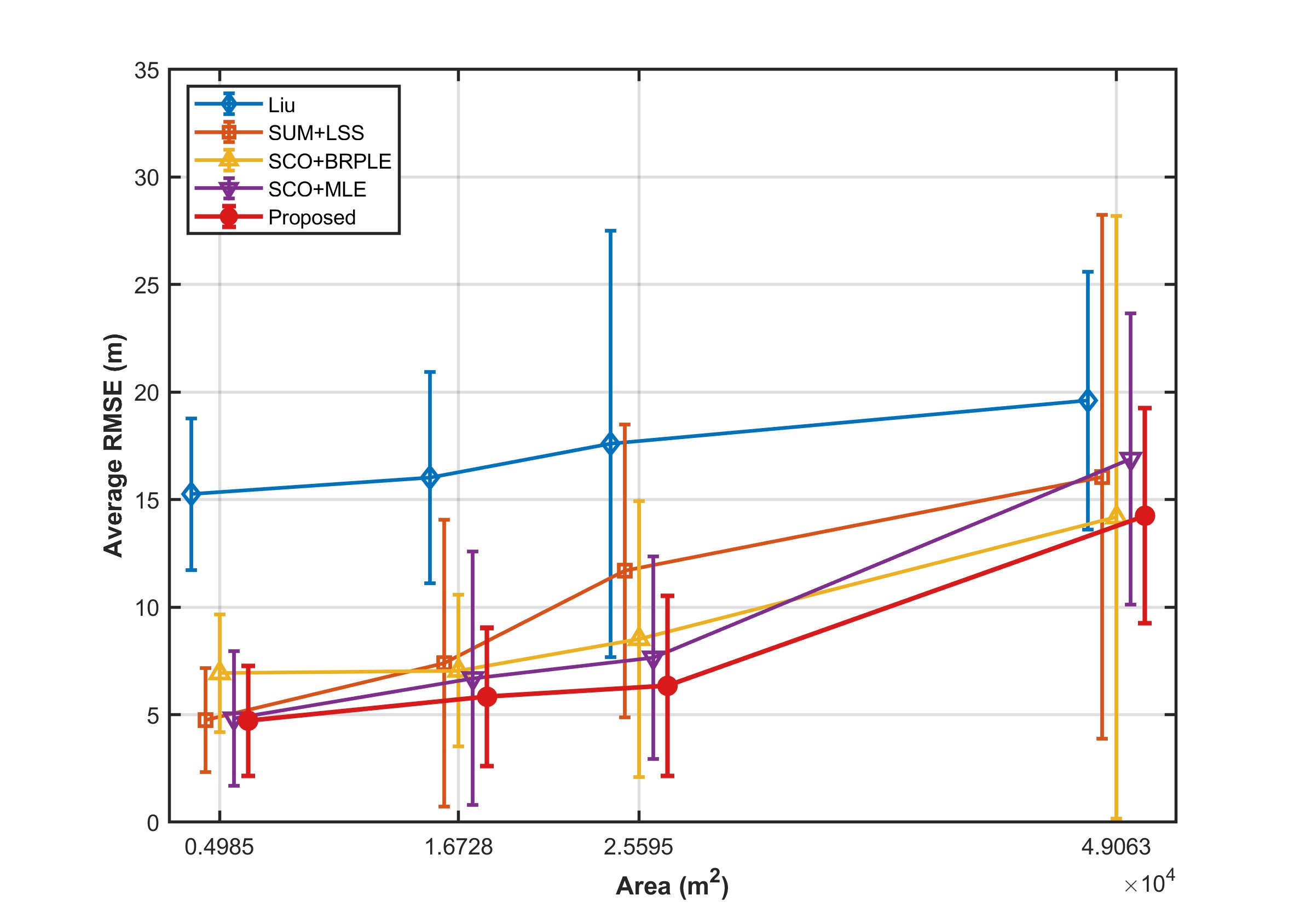}
    \caption{Average RMSE vs. the coverage area. Error bars are shown for all compared methods and indicate the SD of localization errors across repeated trials.}
    \label{fig:position}
\end{figure}

The accuracy of position estimation is evaluated using the RMSE between the estimated and true positions. The proposed method is compared with SUM-LSS~\cite{heydari2023scalable}, Liu's method~\cite{liu2021study}, SCO-BRPLE~\cite{wang2015asymptotically}, and SCO-MLE~\cite{wang2017unified}; the baseline methods use the original SCO-based AoA estimation~\cite{9612590} and their corresponding position solvers. Fig.~\ref{fig:position} compares both the average RMSE and the standard deviation (SD) of localization error. The proposed method achieves the best overall performance across the tested areas, reducing the RMSE by $21.9\%$, $15.1\%$, $13.6\%$, and $54.5\%$ relative to SUM-LSS, SCO-BRPLE, SCO-MLE, and Liu's method, respectively. The gain is more pronounced in the medium-scale deployments: compared with SUM-LSS, the RMSE reduction reaches $21.1\%$ and $45.8\%$ at $16{,}728$ and $25{,}595~\textrm{m}^2$, respectively, while the reduction relative to Liu's method remains above $63\%$ in the three smaller areas. At the largest coverage area, the proposed method remains lower than SUM-LSS, SCO-MLE, and Liu's method by $11.3\%$, $15.6\%$, and $27.3\%$, respectively.

The SDs in Fig.~\ref{fig:position} quantify trial-to-trial variations in localization error. The error-bar comparison shows that the proposed method achieves the lowest overall dispersion across repeated trials, reducing the average localization-error SD by $46.6\%$, $43.9\%$, $27.0\%$, and $38.4\%$ relative to SUM-LSS, SCO-BRPLE, SCO-MLE, and Liu's method, respectively. The stability gain is more pronounced in the medium and large deployments, where the baseline methods exhibit larger trial-to-trial variation and are more sensitive to adverse geometry, signal attenuation, and propagation conditions. Therefore, the proposed method improves both mean localization accuracy and experimental repeatability.

\subsection{Measurement uncertainty budget}

Table~\ref{tab:gum_budget} summarizes the area-dependent uncertainty budget. The resulting $U_{95}$ values are $1.189$, $0.693$, $0.929$, and $2.240$~m for coverage areas of $4{,}985$, $16{,}728$, $25{,}595$, and $49{,}063~\mathrm{m}^2$, respectively. The calculation follows~\cite{GUM2008,shirmohammadi2025uq} and evaluates the uncertainty using four contributions, as described below.

The epistemic Type-A term is $u_{\mathrm{e,A}}=0$~m because the TinyML gate does not enter the coordinate solver. The epistemic Type-B term is $u_{\mathrm{e,B}}=0.165$~m and is estimated from the BRPLE--EV model variation using a rectangular distribution. The aleatoric Type-A term is propagated from repeated localization errors, and its area-dependent values are reported in Table~\ref{tab:gum_budget}. The aleatoric Type-B term is $u_{\mathrm{a,B}}=0.120$~m and is propagated from the rectangular ZED-F9P coordinate bound~\cite{inside_gnss_2023}. Repeated-shot variations arising from TDoA estimation (including sampling quantization), acoustic propagation, and residual angular errors are included in the aleatoric Type-A contribution to avoid double counting.

The four contributions are combined and expanded according to
\begin{equation}\label{eq:gum_combination}
\begin{aligned}
u_\mathrm{c}&=\left(u_{\mathrm{e,A}}^2+u_{\mathrm{e,B}}^2+u_{\mathrm{a,A}}^2+u_{\mathrm{a,B}}^2\right)^{1/2},\\
\nu_\mathrm{eff}&=\frac{u_\mathrm{c}^4}{u_{\mathrm{e,B}}^4/\nu_{\mathrm{e,B}}+u_{\mathrm{a,A}}^4/\nu_{\mathrm{a,A}}},\\
U_{95}&=t_{0.975,\lfloor\nu_\mathrm{eff}\rfloor}u_\mathrm{c}.
\end{aligned}
\end{equation}
Here, $t_{0.975,\nu}$ denotes the 0.975 quantile of the Student-$t$ distribution; its use for a two-sided 95\% coverage interval follows Secs.~G.3.4 and G.4.1 of the GUM~\cite{GUM2008}. Regarding the degrees of freedom, the model-form Type-B term uses $\nu_{\mathrm{e,B}}=1$, whereas the fixed specification-based Type-B bound uses $\nu_{\mathrm{a,B}}=\infty$. The area-dependent $\nu_{\mathrm{a,A}}$ values are obtained from repeated measurements and reported in Table~\ref{tab:gum_budget}. The effective degrees of freedom are rounded down when evaluating the two-sided Student-$t$ quantile. Calculations use the unrounded component estimates, whereas the table displays rounded values.

\begin{table}[t]
\centering
\caption{Area-dependent GUM uncertainty results for the conditional coordinate-localization measurand.}
\label{tab:gum_budget}

\renewcommand{\arraystretch}{1.15}
\resizebox{\columnwidth}{!}{%
\begin{tabular}{c|c|c|c|c|c}
\toprule
\textbf{Area (m$^2$)} & \textbf{$u_{\mathrm{a,A}}$ (m)} & \textbf{$\nu_{\mathrm{a,A}}$} & \textbf{$u_\mathrm{c}$ (m)} & \textbf{$\nu_\mathrm{eff}$} & \textbf{$U_{95}$ (m)} \\
\midrule
$4{,}985$  & 0.548 & 33.18  & 0.584 & 33.84 & 1.189 \\
$16{,}728$ & 0.248 & 86.49  & 0.321 & 13.63 & 0.693 \\
$25{,}595$ & 0.414 & 127.83 & 0.462 & 47.16 & 0.929 \\
$49{,}063$ & 0.674 & 3.07   & 0.704 & 3.62  & 2.240 \\
\bottomrule
\end{tabular}}

\end{table}


\subsection{Computational Complexity}

The computational complexity of the different angle-position estimation methods was evaluated by testing the processing time using a MATLAB 2024b implementation, executed on a PC with an Intel i5-12600k processor and 32 GB of memory. As shown in Table~\ref{proc}, the angle estimation stage accounts for the majority of the localization processing time: SCO-EV and SCO each require 8.6~ms, whereas SUM-LSS requires 7.2~ms. For source position estimation, SCO-EV and SCO-BRPLE each require 0.201~ms, while Liu's method and SUM-LSS require 0.003~ms and 0.002~ms, respectively. In contrast, SCO-MLE requires 112~ms. Thus, SCO-EV provides the improved localization accuracy reported above with the same angle-estimation time as SCO and the same position-estimation time as SCO-BRPLE. These localization times are evaluated separately from the on-device gunshot-recognition processing time reported in Table~\ref{gunshotdetection}.

\begin{table}[ht!]
\centering
\caption{Processing Time for Different Methods}
\begin{tabular}{c|ccc}
\textbf{Angle Est.} & \textbf{SCO-EV} & \textbf{SCO} & \textbf{SUM-LSS} \\
\hline
\textbf{Proc. Time (ms)} & 8.6 & 8.6 & 7.2 \\
\hline
\end{tabular}

\vspace{0.5cm} 

\begin{tabular}{c|ccccc}
\textbf{Position Est.} & \textbf{SCO-EV} & \textbf{Liu's} & \textbf{SUM-LSS} & \textbf{BRPLE} & \textbf{MLE} \\
\hline
\textbf{Proc. Time (ms)} & 0.201 & 0.003 & 0.002 & 0.201 & 112 \\
\hline
\end{tabular}

\label{proc}
\end{table}

Low SNR makes the cross-correlation peak less distinguishable, increasing TDoA uncertainty and localization error. As shown in Fig.~\ref{fig:gcc}, eGCC reaches RMSE $<2\times10^{-5}$~s at 1~dB and RMSE $<0.01$~s at $-5$~dB, but its performance degrades as the received SNR decreases further.

Reverberation can produce secondary correlation peaks and bias delay estimates. eGCC reduces unreliable frequency components, and SDES suppresses repeated echo triggers; however, neither eliminates multipath-induced TDoA bias. Reliable TDoA estimation may fail when strong reflections dominate the direct path.

\section{Conclusion}

This work presents a comprehensive and scalable real-time acoustic surveillance system for gunshot detection and localization, tailored for large-area deployment in anti-poaching applications. The proposed framework integrates cost-effective hardware with lightweight, edge-optimized algorithms, enabling continuous environmental monitoring through multiple autonomous units. Each unit employs a low-power SDES algorithm, followed by a Tiny Machine Learning (TinyML)-based classifier for efficient gunshot identification. Detected gunshot events are locally processed using the eGCC algorithm to estimate TDoA, which is subsequently used by the central server to localize the source via the SCO-EV algorithm. All performance results reported in this work are supported by experimental evaluations. For selected algorithms, simulations are additionally used as auxiliary analyses to more clearly illustrate the respective algorithmic gains under controlled conditions. The proposed approach achieved a MCC of 0.70 for detection reliability and a localization error of 6.34 meters over a $25{,}000\,\mathrm{m}^2$ area. These results demonstrate the practical potential of the proposed system for distributed gunshot surveillance in GNSS-challenged large areas. Future work will focus on improving the robustness of acoustic classification using advanced AI techniques and investigating multimodal sensing approaches by integrating acoustic data with radio frequency and optical signals to further enhance situational awareness and detection accuracy.


\bibliographystyle{IEEEtran}
\bibliography{IEEEabrv.bib,Dengke_bibliography}

@inproceedings{rabinkin1996dsp,
  title={{DSP} implementation of source location using microphone arrays},
  author={Rabinkin, Daniel V and Renomeron, Richard J and Dahl, AJ and French, Joseph C and Flanagan, James L and Bianchi, Mike},
  booktitle={Proc. SPIE Adv. Signal Process. Algorithms, Archit., Implementations VI},
  volume={2846},
  pages={88--99},
  year={1996},
month     = {Aug.}
}

@article{wang2015asymptotically,
  title={An asymptotically efficient estimator in closed-form for 3-{D} {AoA} localization using a sensor network},
  author={Wang, Yue and Ho, KC},
  journal={IEEE Trans. Wireless Commun.},
  volume={14},
  number={12},
  pages={6524--6535},
  year={2015},
month={Jul.},
  publisher={IEEE}
}

@article{liu2021study,
  title={Study on the localization method of multi-aperture acoustic array based on {TDoA}},
  author={Liu, Haitao and Chen, Yonghua and Huang, Yi and Cheng, Xianfu and Xiao, Qian},
 journal   = {IEEE Sens. J.},
  volume={21},
  number={12},
  pages={13805--13814},
  year={2021},
month     = {Jun.},
  publisher={IEEE}
}

@article{1,
  title={4 things we've done to improve animal welfare},
  author={Foster Jacqueline; McClarkin Emma; Flack and John},
  journal={Conservatives in the European Parliament},
  year={2018},
}

@misc{wcsElephants,
  author = {{Wildlife Conservation Society}},
  title = {African Elephants},
  year = {2024},
  url = {https://www.wcs.org/our-work/species/african-elephants},
  note = {Accessed: Apr. 5, 2025}
}

@article{heydari2023scalable,
  title={Scalable real-time sound source localization method based on {TDoA}},
  author={Heydari, Zahra and Mahabadi, Aminollah},
   journal   = {Multimed. Tools Appl.},
  volume={82},
  number={15},
  pages={23333--23372},
  year={2023},
month     = {Aug.},
  publisher={Springer}
}

@misc{fesliyan_studios,
  author = {FesliyanStudios},
  title = {Royalty Free Music and Sound Effect},
  year = {2024},
  url = {https://www.fesliyanstudios.com/},
}

@article{kamminga2018poaching,
  title={Poaching detection technologies—a survey},
  author={Kamminga, Jacob and Ayele, Eyuel and Meratnia, Nirvana and Havinga, Paul},
  journal={Sensors},
  volume={18, no. 5, Art. no. 1474},
  year={2018},
month={May},
  publisher={MDPI}
}

@misc{infineon_datasheet,
  author       = {{Infineon Technologies AG}},
  title        = {{IM72D128} Datasheet},
  year         = {2022},
  url          = {https://www.infineon.com/cms/en/product/sensor/mems-microphones/mems-microphones-for-consumer/im72d128/}
}

@misc{soundthinking,
  author       = {{SoundThinking, Inc.}},
  title        = {SoundThinking: Solutions for Safer Communities},
  year         = {n.d.},
  url          = {https://www.soundthinking.com/},
  note         = {Accessed: 2024-07-15}
}

@article{kotus2014detection,
  title={Detection and localization of selected acoustic events in acoustic field for smart surveillance applications},
  author={Kotus, Jozef and Lopatka, Kuba and Czyzewski, Andrzej},
  journal={Multimed. Tools Appl.},
  volume={68},
  number={1},
  pages={5--21},
  year={2014},
  doi={10.1007/s11042-012-1183-0},
  publisher={Springer}
}

@article{ho2023integrating,
  title={Integrating {AoA} with {TDoA} for joint source and sensor localization},
  author={Ho, KC and Le, Trung-Kien},
  journal={IEEE Trans. Signal Process.},
  volume={71},
  pages={2087--2102},
month={May},
  year={2023},
  publisher={IEEE}
}

@article{sun2024projection,
  title={Projection-Based Algorithm and Performance Analysis for {TDoA} Localization in {MPR}},
  author={Sun, Yimao and Ho, KC and Xing, Tianyi and Yang, Yanbing and Chen, Liangyin},
  journal={IEEE Trans. Signal Process.},
volume={72},
  pages={896--911},
  year={2024},
month={Jan.},
  publisher={IEEE}
}

@inproceedings{zou2020simple,
  title={A simple and efficient iterative method for {ToA} localization},
  author={Zou, Yanbin and Liu, Huaping},
  booktitle={Proc. IEEE Int. Conf. Acoust., Speech Signal Process. (ICASSP)},
  pages={4881--4884},
  year={2020},
month={Apr.},
}

@misc{GSA_47QTCA22D0036_Mod0024_060325,
  author       = {{U.S. General Services Administration}},
  title        = {SIEMENS AUTHORIZED FEDERAL SUPPLY SCHEDULE PRICE LIST},
  year         = {2025},
  url          = {https://www.gsaadvantage.gov}
}

@misc{HEADAcoustics_HEAD_VISOR_VMA_V,
  author       = {{HEAD acoustics GmbH}},
  title        = {Acoustic Camera {HEAD VISOR VMA V}},
  year         = {2025},
  url          = {https://www.head-acoustics.com}
}

@article{qureshi2022gunshots,
  title={Gunshots localization and classification model based on wind noise sensitivity analysis using extreme learning machine},
  author={Qureshi, Shahzad Ahmad and Hussain, Lal and Alshahrani, Haya Mesfer and Abbas, Syed Rahat and Nour, Mohamed K and Fatima, Nayabb and Khalid, Muhammad Imran and Sohail, Huniya and Mohamed, Abdullah and Hilal, Anwer Mustafa},
  journal={IEEE Access},
  month     = {Nov.},
  volume={10},
  pages={87302--87321},
  year={2022},
  publisher={IEEE}
}

@article{chen2024deep,
  title={Deep Transfer Learning Based Intelligent Gunshot Detection and Firearm Recognition Using Tri-Axial Acceleration},
  author={Chen, Zhicong and Zheng, Haoxin and Wu, Lijun and Huang, Jingcang and Yang, Yang},
  journal={IEEE Internet Things J.},
 year={2025},
  volume={12},
  number={5},
month={Nov.},
  pages={5891-5900}
}

@book{lyons1997understanding,
  title={Understanding digital signal processing, 3/E},
  author={Lyons, Richard G},
  year={1997},
month     = {Jan.},
  publisher={Pearson Education India}
}

@misc{soundbible,
  author       = {{SoundBible}},
  title        = {SoundBible: Free Sound Clips for Download},
  year         = {2024},
  url          = {https://soundbible.com/},
  note         = {Accessed: 2024-12-20}
}

@misc{freesound,
  author       = {{Freesound}},
  title        = {Freesound: Collaborative Database of Audio Samples},
  year         = {2024},
  url          = {https://freesound.org/},
  note         = {Accessed: 2024-12-20}
}

@inproceedings{svatos2019smart,
  title={Smart acoustic sensor},
  author={Svatos, Jakub and Holub, Jan},
  booktitle={Proc. IEEE Int. Forum Res. Technol. Soc. Ind. (RTSI)},
  pages={161--165},
  year={2019},
month     = {Oct.},
}

@ARTICLE{9612590,
  author={Sun, Yimao and Ho, K. C. and Wang, Gang and Chen, Hongyang and Yang, Yanbing and Chen, Liangyin and Wan, Qun},
  title={Computationally Attractive and Location Robust Estimator for {IoT} Device Positioning}, 
   journal={IEEE Internet Things J.},
  volume={9},
  number={13},
  pages={10891--10907},
  year={2021},
month={Nov.},
  publisher={IEEE}
}

@article{sun2018solution,
  title={Solution and analysis of {TDoA} localization of a near or distant source in closed form},
  author={Sun, Yimao and Ho, KC and Wan, Qun},
  journal={IEEE Trans. Signal Process.},
  volume={67},
  number={2},
  pages={320--335},
  year={2018},
month={Nov.},
  publisher={IEEE}
}

@article{makinen2010shooter,
  title={Shooter localization and bullet trajectory, caliber, and speed estimation based on detected firing sounds},
  author={M{\"a}kinen, Toni and Pertil{\"a}, Pasi},
  journal={Appl. Acoust.},
  volume={71},
  number={10},
  pages={902--913},
  year={2010},
month     = {Oct.},
  publisher={Elsevier}
}

@conference{hrabina2018gunshot,
  title={Gunshot recognition using low level features in the time domain},
  author={Hrabina, Martin and Sigmund, Milan},
  booktitle={Proc. 2018 28th Int. Conf. Radioelektronika},
  pages={1--5},
  month     = {Apr.},
  year={2018},
  doi={10.1109/RADIOELEK.2018.8376372}
}

@inproceedings{ahmed2013improving,
  title={Improving efficiency and reliability of gunshot detection systems},
  author={Ahmed, Talal and Uppal, Momin and Muhammad, Abubakr},
  booktitle = {Proc. IEEE Int. Conf. Acoust., Speech Signal Process. (ICASSP)},
month     = {May},
  year      = {2013},
  pages     = {513--517}
}

@article{roma2018environmental,
  title={Environmental sound recognition using short-time feature aggregation},
  author={Roma, Gerard and Herrera, Perfecto and Nogueira, Waldo},
  journal={J. Intell. Inf. Syst.},
  volume={51},
  number={3},
  month     = {Jun.},
  pages={457--475},
  year={2018},
  publisher={Springer}
}

@article{wu2024multi,
  title={Multi camera localization handover based on {YOLO} object detection algorithm in complex environments},
  author={Wu, Wennan and Lai, Jizhou},
  journal={IEEE Access},
  volume={12},
  pages={15236--15250},
  year={2024},
  month     = {Jan.},
  publisher={IEEE}
}

@misc{inside_gnss_2023,
  author       = {{u-blox}},
  title        = {{ZED-F9P} Module},
  year         = {2024},
  url          = {https://www.u-blox.com/en/product/zed-f9p-module},
  note         = {Accessed: 2024-12-20}
}

@misc{sony_spresense,
  author       = {{Sony Developer World}},
  title        = {Spresense Development Guides - Introduction},
  year         = {2024},
  url          = {https://developer.sony.com/spresense/development-guides/introduction_en.html},
  note         = {Accessed: 2024-12-20}
}

@misc{Xinxiwang,
  author       = {{Xinxiwang}},
  title        = {{100-W} High-Power Handheld Loudspeaker Outdoor Loudspeaker Amplifier},
  year         = {2024},
  month        = {Dec.},
  url          = {https://item.taobao.com/item.htm?_u=lnt7i8f5a4c&id=548662766724&spm=a1z09.2.0.0.65eb2e8dKkmAZZ},
  note         = {Accessed: Aug. 6, 2026}
}

@article{alimi2023situ,
  title={In situ underwater localization of magnetic sensors using natural computing algorithms},
  author={Alimi, Roger and Fisher, Elad and Nahir, Kanna},
  journal={Sensors},
  volume={23, no. 4, Art. no. 1797},
  year={2023},
  month     = {Feb.},
  publisher={MDPI}
}

@article{wang2022optimization,
  title={Optimization Algorithm for Delay Estimation Based on Singular Value Decomposition and Improved {GCC-PHAT} Weighting},
  author={Wang, Shizhe and Li, Zongji and Wang, Pingbo and Chen, Huadong},
  journal={Sensors},
  volume={22, no. 19, Art. no. 7254},
  year={2022},
month     = {Oct.},
  publisher={MDPI}
}

@article{mozer2023introduction,
  title={An introduction to illegal wildlife trade and its effects on biodiversity and society},
  author={Mozer, Annika and Prost, Stefan},
  journal={Forensic Sci. Int.: Animals Environ.},
  volume={3},
  pages={1--12},
month={Dec.},
  year={2023},
  publisher={Elsevier}
}

@article{kumar2023survey,
  title={A survey on localization for autonomous vehicles},
  author={Kumar, Debasis and Muhammad, Naveed},
  journal={IEEE Access},
 volume={11},
  pages={115865--115883},
  year={2023},
month={Oct.},
  publisher={IEEE}
}

@inproceedings{kang2016pre,
  title={On pre-filtering strategies for the {GCC-PHAT} algorithm},
  author={Kang, Hong-Goo and Graczyk, Michael and Skoglund, Jan},
 booktitle = {Proc. IEEE Int. Workshop Acoust. Signal Enhanc.},
  pages     = {1--5},
  month     = {Sep.},
  year      = {2016}
}

@inproceedings{hymel2022edge,
  title={Edge Impulse: An {MLOps} Platform for Tiny Machine Learning},
  author={Hymel, Shawn and others},
  booktitle={Proc. Mach. Learn. Syst. (MLSys)},
  volume={5},
  year={2023}
}

@inproceedings{morehead2019low,
  title={Low cost gunshot detection using deep learning on the raspberry pi},
  author={Morehead, Alex and Ogden, Lauren and Magee, Gabe and Hosler, Ryan and White, Bruce and Mohler, George},
 booktitle = {Proc. 2019 IEEE Int. Conf. Big Data (Big Data)},
  year      = {2019},
 month     = {Dec.},
  pages     = {3038--3044}
}

@article{zhang2014wildlife,
  title={A wildlife monitoring system based on wireless image sensor networks},
  author={Zhang, Junguo and Luo, Xin and Chen, Chen and Liu, Zhen and Cao, Shuai},
  journal={Sens. Transducers},
  volume={180},
  number={10},
  pages={104--109},
  year={2014},
  month     = {Oct.},
  publisher={IFSA Publishing, SL}
}

@techreport{GUM2008,
  author      = {{Joint Committee for Guides in Metrology}},
  title       = {Evaluation of measurement data---Guide to the expression of uncertainty in measurement},
  institution = {Bureau International des Poids et Mesures},
  type        = {JCGM 100:2008(E)},
  year        = {2008},
  url         = {https://doi.org/10.59161/JCGM100-2008E}
}

@article{shirmohammadi2025uq,
  title={Review and Performance Evaluation of Uncertainty Quantification in Data-Driven {AI}-Assisted Measurements},
  author={Shirmohammadi, Shervin and Wang, Fan and Hsu, Cheng-Hsin},
  journal={IEEE Open J. Instrum. Meas.},
  volume={4},
  note={{Art. no. 2500722}},
  year={2025},
  month={Oct.},
  publisher={IEEE},
  doi={10.1109/OJIM.2025.3621742}
}

@article{zengin2024hybrid,
  title={A Hybrid Model for 3-{D} Gunshot Localization Using Muzzle Blast Sound Only},
  author={Zengin, Kaz{\i}m and Ye{\c{s}}ildirek, Ayd{\i}n},
  journal={IEEE Trans. Instrum. Meas.},
  volume={73, Art. no. 2533810},
  year={2024},
month={Oct.},
  publisher={IEEE}
}

@article{vakulya2011fast,
  title={Fast Adaptive Acoustic Localization for Sensor Networks},
  author={Vakulya, Gergely and Simon, Gyula},
  journal={IEEE Trans. Instrum. Meas.},
  volume={60},
  number={5},
  pages={1820--1829},
  year={2011},
month={May},
  publisher={IEEE}
}

@article{shen2024fourth,
  title={Fourth-Order Cumulant-Based Acoustic Imaging via Small-Aperture Acoustic Vector-Sensor Array},
  author={Shen, Xingchen and Bao, Ming and Chen, Zhifei and Zhao, Jing and Gao, Yan},
  journal={IEEE Trans. Instrum. Meas.},
  volume={73},
  pages={1--12},
  year={2024},
month={Jun.},
  publisher={IEEE}
}

@article{bharathi2019time,
  title={Time delay estimation in reverberant and low {SNR} environment by {EMD} based maximum likelihood method},
  author={Bharathi, B. Marxim Rahula and Mohanty, A. R.},
  journal={Measurement},
  volume={137},
  pages={655--663},
  year={2019},
month={Apr.},
  publisher={Elsevier}
}

@article{jeong2026micgraphnet,
  title={{MicGraphNet}: Microphone graph network for cross-correlation-based time delay estimation for accurate sound source localization},
  author={Jeong, Iljoo and Huh, Hyunsuk and Park, Bumsoo and Lee, Anna and Jung, In-Jee and Lee, Seungchul},
  journal={Measurement},
  volume={258, Art. no. 119552},
  year={2026},
month={Jan.},
  publisher={Elsevier}
}

@article{svatos2023impulse,
  title={Impulse acoustic event detection, classification, and localization system},
  author={Svatos, Jakub and Holub, Jan},
  journal={IEEE Trans. Instrum. Meas.},
  volume={72, Art. no. 6501515},
  year={2023},
month     = {Feb.},
  publisher={IEEE}
}

@article{abiri2020bullet,
  title={The bullet shockwave-based real-time sniper sound source localization},
  author={Abiri, Aghil and Parsayan, Ali},
  journal={IEEE Sens. J.},
  volume={20},
  number={13},
month={Jul.},
  pages={7253--7264},
  year={2020},
  publisher={IEEE}
}

@article{sallai2011weapon,
  title={Weapon classification and shooter localization using distributed multichannel acoustic sensors},
  author={Sallai, Janos and Hedgecock, Will and Volgyesi, Peter and Nadas, Andras and Balogh, Gyorgy and Ledeczi, Akos},
  journal={J. Syst. Archit.},
  volume    = {57},
  number    = {10},
  pages     = {869--885},
  month     = {Oct.},
  year      = {2011}
}

@inproceedings{manring2023low,
  title={Low-Cost Gunshot Detection System with Localization for Community Based Violence Interruption},
  author={Manring, Isaac and Hill, James H and Mohler, George and Brantingham, P Jeffrey and Williams, Thomas and White, Bruce},
  booktitle = {Proc. IEEE 10th Int. Conf. Data Sci. Adv. Anal.},
  pages={1--7},
  month     = {Oct.},
  year={2023},
}

@ARTICLE{9040635,
  author={Sun, Yimao and Ho, K. C. and Wan, Qun},
  journal={IEEE Trans. Signal Process.}, 
  title={Eigenspace Solution for {AoA} Localization in Modified Polar Representation}, 
  year={2020},
month     = {Apr.},
  volume={68},
  number={},
  pages={2256-2271},
  doi={10.1109/TSP.2020.2981773}}

@ARTICLE{9166461,
  author={Sanchez-Iborra, Ramon and Skarmeta, Antonio F.},
  journal={IEEE Circuits Syst. Mag.}, 
  title={{TinyML}-Enabled Frugal Smart Objects: Challenges and Opportunities}, 
  year={2020},
month={Aug.},
  volume={20},
  number={3},
  pages={4-18},
  doi={10.1109/MCAS.2020.3005467}}

@article{wang2017unified,
  title={Unified near-field and far-field localization for {AoA} and hybrid {AoA}-{TDoA} positionings},
  author={Wang, Yue and Ho, K. C.},
  journal={IEEE Trans. Wireless Commun.},
  volume={17},
  number={2},
  pages={1242--1254},
  year={2018},
  month={Feb.},
  doi={10.1109/TWC.2017.2777457}
}

@article{de2021timely,
  title={Timely poacher detection and localization using sentinel animal movement},
  author={de Knegt, Henrik J and Eikelboom, Jasper AJ and van Langevelde, Frank and Spruyt, W Fran{\c{c}}ois and Prins, Herbert HT},
  journal={Sci. Rep.},
  volume={11},
  number={1},
  pages={1--11},
  year={2021},
month={Feb.},
  publisher={Nature Publishing Group UK London}
}

@article{burgues2020environmental,
  title={Environmental chemical sensing using small drones: A review},
  author={Burgu{\'e}s, Javier and Marco, Santiago},
  journal={Sci. Total Environ.},
  volume={748},
  pages={1--20},
  year={2020},
  month     = {Dec.},
  publisher={Elsevier}
}

@article{choudhary2023survey,
  title={A survey on seismic sensor based target detection, localization, identification, and activity recognition},
  author={Choudhary, Priyankar and Goel, Neeraj and Saini, Mukesh},
  journal={ACM Comput. Surv.},
  volume={55},
  number={11},
  pages={1--36},
  year={2023},
  month     = {Nov.},
  publisher={ACM New York, NY}
}

@article{matata2022electrified,
  title={Electrified fencing as a mitigation strategy for human-elephant conflict in Western Serengeti: Community perspectives},
  author={Matata, Maria T and Kegamba, Juma J and Mremi, Rudolf and Eustace, Abraham},
 journal   = {J. Nat. Conserv.},
  volume    = {70},
  pages     = {1--10},
  month     = {Aug.},
  year      = {2022},
  publisher={Elsevier}
}

@article{zhu2023coaxial,
  title={Coaxial cable sensing: Review and perspective},
  author={Zhu, Chen and Huang, Jie},
  journal={IEEE Trans. Microw. Theory Techn.},
  volume={72},
  number={3},
  pages={1790--1809},
  year={2023},
month={Aug.},
  publisher={IEEE}
}

@IEEEtranBSTCTL{IEEEexample:BSTcontrol,
  CTLdash_repeated_names = {no}
}
\end{document}